\documentclass[12pt,a4paper]{article}
\usepackage{amsmath}
\usepackage{amssymb}
\usepackage{amsthm}
\usepackage{float}
\usepackage{amsfonts}
\usepackage{graphicx}
\usepackage[]{cite}
\usepackage{float}
\usepackage{subfigure}
\usepackage{verbatim}
\usepackage[left=2cm,right=2cm,top=3cm,bottom=2.5cm]{geometry}
\usepackage[numbers]{natbib}
\usepackage[utf8]{inputenc}
\def\case#1/#2{\frac{#1}{#2}}

\def \D {\tilde{\nabla}}
\def\la {\langle}
\def\ra {\rangle}
\newcommand{\sfrac}[2]{{\textstyle{#1\over#2}}}
\def \ep {\varepsilon}
\def\tl{\tilde}
\def\rd {\displaystyle{\cdot}}
\def\ts {\textstyle}

\usepackage[usenames,dvipsnames,svgnames]{xcolor}
\usepackage[colorlinks=true,
      linkcolor=red,
      urlcolor=gray,
      citecolor=blue]{hyperref}

\def\myalign#1{%
  \def\trule{\noalign{\smallskip\hrule\medskip}}
  \def\nebc{\nearrow\bigcup}
  \def\sebc{\searrow\bigcup}
  \def\pminf{{}_{-\infty}|^{+\infty}}
  \let\Inf\infty
  \def\amp{&} 
  \vbox{\mathsurround0pt\openup1\jot
    \halign{%
      &$\displaystyle##\hfil\tabskip0pt$&\amp##\tabskip1em\crcr
      \noalign{\hrule height1pt\smallskip}#1\noalign{\smallskip\hrule height1pt}\crcr}}}

\begin{document}
\begin{center}
\textbf{On covariant perturbations with scalar field in modified Gauss-Bonnet gravity}
\end{center}
\hfill\\
Albert Munyeshyaka$^{1}$,Joseph Ntahompagaze$^{2}$,Tom Mutabazi$^{1}$ and Manasse.R  Mbonye$^{2,3,4}$\\
\hfill\\ 
$^{1}$Department of Physics, Mbarara University of Science and Technology, Mbarara, Uganda\;\;\; \; \;\hfill\\
$^{2}$Department of Physics, College of Science and Technology, University of Rwanda, Rwanda\;\;\; \; \;\hfill\\ 
$^{3}$ International Center for Theoretical Physics (ICTP)-East African Institute for Fundamental Research, University of Rwanda, Kigali, Rwanda
\;\;\; \; \;\hfill\\
$^{4}$ Rochester Institute of Technology, NY, USA.\\ \\\\
Correspondence:munalph@gmail.com\;\;\;\;\;\;\;\;\;\;\;\;\;\;\;\;\;\;\;\;\;\;\;\;\;\;\;\;\;\;\;\;\;\;\;\;\;\;\;\;\;\;\;\;\;\;\;\;\;\;\;\;\;\;\;\;\;\;\;\;\;\;\;\;\;\;\;\;\;\;\;\;\;\;\;
\begin{center}
\textbf{Abstract}
\end{center}
We investigate  cosmological perturbations of $f(G)$ gravity in the presence of a scalar field.  Using the $1+3$ covariant formalism, we present the energy overdensity perturbation equations responsible for large scale structure formation. After applying harmonic decomposition method together with the redshift transformation technique, we obtain the fully perturbed equations in redshift space. The equations are solved to  study  the growth of matter overdensities contrast with redshift. For both short- and long-wavelength modes, we obtain numerical results for  particular functional form $f(G)$ models and scalar field. We find that, for this choice the energy overdensity perturbations decay with increase in redshift.  However, for both  short- and long-wavelength modes, the perturbations which include amplitude effects due to the $f(G)$ models with a scalar field do differ remarkably from those in $\Lambda$CDM. The results reduce to GR results in the limit of $f(G)\rightarrow G$ and in the absence of scalar field.
\\
\hfill\\
\textit{keywords:} $1+3$ covariant formalism-- cosmic acceleration-- Cosmological perturbations-- scalar-field.\\
\textit{PACS numbers:} 04.50.Kd, 98.80.-k, 95.36.+x, 98.80.Cq; MSC numbers: 83F05, 83D05.
(This Article was accepted for publication in European Physical Journal C on December 22, 2023)
\section{Introduction}\label{introduction}
Recent observational studies such as high redshift supernovae $I_{a}$ \cite{filippenko1998results,perlmutter1999measurements,riess2004type}, cosmic microwave background (CMB)  \cite{sherwin2011evidence,das2011detection} and baryon acoustic oscillations (BAO) \cite{eisenstein2005detection} have provided convincing evidence for an  accelerated expansion of the Universe.  The standard model of cosmology referred to as  $\Lambda$CDM  which includes the  cosmological constant  introduces inflation addresses several problems in the early universe, including  the horizon problem, the flatness problem and the monopole problem, the model is also consistent with various astrophysical and cosmological observations such as the origin of cosmic microwave background radiations (CMB), the formation and distribution of large scale structure, the synthesis of the light elements in the universe and its expansion \cite{riess1998observational, caldwell2009physics, silvestri2009approaches,weinberg2013observational}. However this $\Lambda$CDM model currently shows  limitations in explaining some cosmological issues in both the early and late time Universe. These include  the cosmological constant problem, the coincidence problem, the fine tunning problem, and the current problems of Hubble tensions and Sigma $8$. The model still cannot explain the nature and even the source  of both dark  matter and dark energy problems. As a result, these shortcomings have encouraged a search for alternative approaches  seeking to address some of these problems. These include the use of  different scalar fields to describe the cosmic acceleration, they include the consideration of Chaplygin gas models \cite{swart2019unifying,sahlu2019chaplygin,gadbail2022generalized,twagirayezu2023chaplygin,azizi2017reconstruction,hough2021confronting,saadat2013viscous,elmardi2016chaplygin,paul2013observational}. On the other hand, these latter models present severe constraints on  the matter power spectrum in galaxy clusters.\\ Another  alternative is to modify gravity on large scale. Several different modified theories of gravity which are  extensions of Einstein's gravity  in the past have been introduced to address early and late time cosmic problems. They include Scalar-Tensor theories, $f(R)$ models, $f(T)$ models and $f(G)$ models of gravity, to name but a few, where $R$, $T$ and $G$ are the Ricci scalar, torsion tensor and Gauss-Bonnet invariant, respectively  \cite{li2007cosmology,nojiri2008modified,makarenko2016role,makarenko2017asymptotic,sami2017inflationary,nojiri2017modified,ntahompagaze2017f,odintsov2017unification,sami2018reconstructing,capozziello2019cosmological,odintsov2019unification,nojiri2020unifying,odintsov2020geometric,venikoudis2022late}. These theories are capable of unifying  the  early inflationary era and the late-time era in a way similar to the  $\Lambda$CDM model. It has been suggested that some of these models may have  issues to address such as achieving a consistent description of neutron stars, as pointed out in \cite{kobayashi2009can,babichev2010relativistic}, dealing with  singularity problems, as pointed out in  \cite{bamba2008future,bamba2010finite}, controlling matter instability, as pointed out in  \cite{dolgov2003can}, and satisfying  solar system tests, as pointed out in  \cite{nojiri2008inflation,nojiri2006dark,bamba2010finite}. In some of these models such as $\frac{1}{R}$ model, it was found \cite{nojiri2005modified,bamba2008future} that the Newtonian limit may not be recovered. For several of these $\frac{1}{R}$ models, the corrections to Newton's law do not comply with  the solar system tests.\\ \\Most of these challenges appear to be in absence in $f(G)$ as investigated in \cite{nojiri2011unified}. As a result, there has been  a marked interest in the use of modified $f(G)$ models. For examples, It was demonstrated that scalar field coupled with Gauss-Bonnet invariant and with a potential has no ghosts  and is stable \cite{nojiri2011unified,nojiri2005gauss,calcagni2005dark}.\\  The work conducted by Nojiri et al.  \cite{nojiri2011unified} investigated the modified Gauss-Bonnet gravity and demonstrated that it may lead to a unified cosmic history. Further, the authors in \cite{cognola2007string,nojiri2007dark}  demostrated that some scalar-Gauss-Bonnet gravities may be compatible with the known classical history of the universe expansion (radiation, matter dominance, transition to deceleration and acceleration).\\ In \cite{nojiri2007dark}, the authors proposed a unified $f(R)$-scalar-Gauss-Bonnet gravity for dark energy models. A reconstruction program for such models was later developed. It was further shown that Gauss-Bonnet term may play an important role in another class of gravitational models where it couples with scalar field kinetic term.
\\ \\  In \cite{odintsov2020unification}, the authors focussed mainly on a model of the form $R+f(G)$ in order to avoid the presence of primordial superluminal perturbation modes. That work finds an appropriate model of Gauss-Bonnet gravity which may describe the dark energy era,it also demonstrated that in some Gauss-Bonnet models, it is possible to provide a unified description of inflationary  and dark energy era with the same model.\\\\
 In the work done by \cite{oikonomou2016gauss,venikoudis2022late}, the authors presented the phenomenology of the late time universe of a given scalar-tensor theory in the context of string inspired gravity.  This theory involves a scalar field which is minimally coupled with the Ricci scalar and a function of $f(G)$ \cite{nojiri2005gauss,nojiri2005modified,nojiri2006dark,oikonomou2020non,odintsov2020rectifying,oikonomou2021refined,odintsov2021canonical}. Here the coupling between the scalar field and the Gauss-Bonnet invariant was ignored for simplicity reasons. The combination of both a scalar field and  $f(G)$ models has the advantage that a unified description between early and late time epochs may be achieved. One possible scenario involves a dominant scalar field during the inflationary era while in the late time era, the curvature corrections in the Gauss-Bonnet invariant can drive the accelerated expansion of the universe. The validity of these theories has been studied and can provide the viable phenomenological models in the primordial late time era \cite{fronimos2021inflation}.
    There is a need to study whether  this combination at perturbative level  can produce  significant results for large scale structure formation and can be treated as the form of dark energy to source cosmic acceleration. This can best be approached through  the study of $1+3$ covariant perturbations using the combination of both scalar field and Gauss-Bonnet $f(G)$ model.\\ \\  There are two approaches to study cosmological perturbations namely metric based approach \cite{bardeen1980gauge,dunsby1992covariant,perico2017running,borges2020growth,sharma2021growth} and the $1+3$ covariant approach \cite{ellis1989covariant,dunsby1992cosmological,bruni1992gauge,sahlu2020scalar,ntahompagaze2020multifluid}. In the latter approach, the perturbations defined describe the true physical degrees of freedom and there are no unphysical modes present. This approach has been used to study the cosmological perturbations in different contexts of general relativity and different alternative theories of gravity \cite{maartens1998covariant,sami2021perturbations}. For instance the work done in \cite{sami2021perturbations} considered the covariant form of the field equations of $f(R)$ gravity as a subclass of scalar tensor to study the linear cosmological perturbations.  In \cite{borges2008evolution} they analysed the evolution of density perturbations in a particular spatially flat cosmological model with a vacuum decay. In \cite{li2007cosmology}, the authors considered the perturbation dynamics in modified Gauss-Bonnet gravity theory at first-order and found that  at perturbation level, the small scale dark matter density perturbations grow much more quickly than in the $\Lambda$CDM paradigm, which might lead to a strong scale dependent matter power spectrum.\\\\
     In the previous research works \cite{munyeshyaka2021cosmological,munyeshyaka20231+,munyeshyaka2023multifluid,munyeshyaka2023perturbations,twagirayezu2023chaplygin}, we applied $1+3$ covariant perturbation in the context of $f(G)$ gravity, and we found that the energy overdensity perturbations decay with increase in redshift.
In the present work, we apply the $1+3$ covariant approach to a mixture of the scalar field and  Gauss-Bonnet fluids. We consider the fluids to be  non-interacting and  study the perturbation of gradient variables of scalar field and Gauss-Bonnet fluids in addition to the gradient variables of physical standard matter in both short and long wavelength modes. \\
In this context, the consideration of two different cases is made while analysing the energy density perturbations. In the first case, we solve the whole system of energy density perturbation equations to analyse the large scale structure formation scenario, where the matter perturbation equations couple with the perturbation equations for scalar field and Gauss-Bonnet energy densities. For comparison purposes, we show that the energy density perturbations decay with increase in redshift for both GR and the considered scalar field assisted $f(G)$ gravity model. In the second case, we consider the quasi-static approximation technique to study the implication of perturbation equations on small scale structure. In this approximation, we assume the very slow fluctuations in the scalar field and Gauss-Bonnet energy densities and their momenta, compared to matter energy density. As a result, the matter energy density decouple from the energy densities of the scalar field and Gauss-Bonnet fluids. \\\\
The rest of this paper is organised as follows: in Sec. (\ref{sec2}), a review of cosmic dynamics in GR is presented; in Sec. (\ref{sec3}),  we  present the $1+3$ covariant form of the field equations of the scalar field assisted $f(G)$ gravity; in Sec. (\ref{sec4})  we define the fluid gradient variables and presents their evolution equations.  We analyse the perturbation equations for  matter fluctuations for both GR and  $f(G)$-scalar field models in Sec. (\ref{sec5}) and Sec. (\ref{sec6}), respectively.  Sec. (\ref{sec7}) gives closing remarks.

Natural units in which $c=8\pi G=1$
will be used throughout this paper, and Greek indices run from 0 to 3.
The symbols $\nabla$, $\D$ and the overdot $^{.}$ represent the usual covariant derivative, the spatial covariant derivative, and differentiation with respect to cosmic time, respectively. The $(-+++)$ spacetime signature will be considered.
\section{Cosmic dynamics-evolution equations in GR}\label{sec2}
On the large scale structure of the universe, the homogeneous and isotropic assumptions imply that the current universe is close to a flat geometry with radius of curvature $R$, same at every point in space and that the universe expansion has to be the same at every space with the scale factor $a(t)$. The relationship betwween curvature and matter content of the universe is given by the Einstein's equation represented as
\begin{eqnarray}
 G_{\mu\nu} \equiv R_{\mu\nu}-\frac{1}{2}g_{\mu\nu}R=8\pi G_{N}T_{\mu\nu},\label{eq1}
\end{eqnarray}
where $G_{\mu\nu}$ is the Einstein tensor, $R_{\mu\nu}$, $R=g^{\mu\nu}R_{\mu\nu}$, $T_{\mu\nu}$ and $G_{N}$ are Ricci tensor, Ricci scalar, energy momentum tensor and Newton gravitational constant respectively and $g^{\mu\nu}$ is the metric tensor. For a perfect fluid, the energy momentum tensor is given by
\begin{eqnarray}
 T_{\mu\nu}=(\rho+p)u^{\mu}u^{\nu}+pg^{\mu\nu},
\end{eqnarray}
where, $\rho$ and $p$ are the energy density and isotropic pressure respectively and $u^{\mu}$ is the $4$-velocity. The Friedmann-Robertson-Walker (FRW) metric is given by \begin{eqnarray}
 ds^{2}=-dt^{2}+a^{2}(t)\Big(dx^{2}+dy^{2}+dz^{2}\Big),
 \label{eq3}
\end{eqnarray}
where $a(t)$ is the scale factor governing the expansion of the Universe and it is associated with the dynamics of the universe. For the metric of the form eq. (\ref{eq3}) and considering a perfect fluid in a flat goemetry, the Friedmann,  acceleration and the continuity equations are represented as
\begin{eqnarray}
 &&H^{2}\equiv\Big(\frac{\dot{a}}{a}\Big)^{2}=\frac{8\pi G_{N}\rho}{3},\label{eq4}\\
 &&\dot{H}=-4\pi G_{N}(\rho+p),\label{eq5}\\
 &&\frac{\ddot{a}}{a}=\frac{-4\pi G_{N}}{3}(\rho+p)\label{eq6},\\
 &&\dot{\rho}+3H(\rho+p)=0,\label{eq7}
\end{eqnarray}
 where $H$ is the Hubble parameter. The eq. (\ref{eq4}) tells us that the universe containing matter has to be dynamically evolving. Knowing that the critical density of matter is given by $\rho_{c}\equiv \frac{3H^{2}}{8\pi G_{N}}$ and the density parameter $\Omega_{i}=\frac{\rho_{i}}{\rho_{c}}$, eq. (\ref{eq4}) can be rewritten as
\begin{eqnarray}
\sum_{i} \Omega_{i}=1,
\end{eqnarray}
where $\Omega_{i}$ is the density parameter of the matter species present in the universe. For a barotropic perfect fluid with an equation of state parameter given by $w=\frac{p}{\rho}$, and from eq. (\ref{eq4}) and eq. (\ref{eq5}), one gets \cite{copeland2006dynamics}
\begin{eqnarray}
 &&H=\frac{2}{3(1+w)(t-t_{0})},\\
 &&a(t)\propto (t-t_{0})^{\frac{2}{3(1+w)}},\\
 &&\rho \propto a^{-3(1+w)},
\end{eqnarray}
whereby for a universe dominated by a dust with an equation of state parameter ($w=0$), or radiation ($w=\frac{1}{3}$), yields $a(t)\propto (t-t_{0})^{\frac{2}{3}}$, $\rho\propto a^{-3}$ and $a(t)\propto (t-t_{0})^{\frac{1}{2}}$, $\rho\propto a^{-4}$ respectively, leading to a decelerated expansion of the universe as presented in eq. (\ref{eq6}). The accelerated expansion ($a(t)>0$) occus for $w<-\frac{1}{3}$.
\section{The $1+3$ covariant form of the field equations of the scalar field assisted $f(G)$ gravity}\label{sec3}
The gravitational action involving a scalar field assisted by $f(G)$ gravity is  presented as \cite{li2007cosmology,venikoudis2022late,azizi2017reconstruction,odintsov2021late,copeland2006dynamics}
\begin{eqnarray}
 S=\int d^{4}x\sqrt{-g}\Big(\frac{R}{2\kappa^{2}}-\frac{1}{2}\bigtriangledown_{\mu}\phi \bigtriangledown^{\mu} \phi-V(\phi)+\frac{f(G)}{2}+\mathcal{L}_{m}\Big)\;
 \label{eq12},
\end{eqnarray}
where $f(G)$ represents an arbitrary function depending on the Gauss-Bonnet invariant $G$ given by  $G=R^{2}-4R_{\mu \nu}R^{\mu \nu}+R_{\mu\nu\alpha \beta} R^{\mu\nu\alpha \beta}$,  where $R^{\mu\nu\alpha \beta}$ is the  Riemann curvature tensor, $g$ is the determinant $g^{\mu\nu}$ and $\kappa$ is the gravitational constant. $ \frac{1}{2}\bigtriangledown_{\mu}\phi \bigtriangledown^{\mu} \phi$ and $V(\phi)$ represent the kinetic term  of the scalar field and the scalar potential respectively.
For the case $f(G)=G$, $\int d^{4}x\sqrt{-g}G=0$, and there is  no scalar field consideration, we recover the  gravitational action for GR, $S=\int d^{4}x\sqrt{-g}\Big(\frac{R}{2\kappa^{2}}+\mathcal{L}_{m}\Big)$. In this context, eq. (\ref{eq12}) produces the field equations represented as
\begin{eqnarray}
 G_{\mu\nu} \equiv R_{\mu\nu}-\frac{1}{2}g_{\mu\nu}R=8\pi G_{N}T^{tot}_{\mu\nu},\label{eq13}
\end{eqnarray}
where $T^{tot}_{\mu\nu}$ represents the energy momentum tensor of the total fluids.
In this paper, we use the $1+3$ covariant formalism, where the spacetime is split into temporal and spatial components with respect to the congruence. This formalism treats the slicing of four dimensional space by time and hypersurface. For congruence normal to a spacelike hypersurface, the $1+3$ covariant formalism reduces to $3+1$ \cite{gourgoulhon20073+}. The basic structure of the $1+3$ covariant formalism is a congruence of one dimensional curves, mostly timelike curves \cite{park2018covariant,gourgoulhon20073+,munyeshyaka2023multifluid}. In this case, we decompose spacetime cosmological manifold into time and space submanifold separately with a perpendicular $4$-velocity field vector $u^{a}$ so that
\begin{eqnarray}
 u^{a}=\frac{dx^{a}}{d\tau}\;,
\end{eqnarray}
with the metric tensor  related to the spatial component as
\begin{eqnarray}
 g_{\mu\nu}=h_{ab}-u_{a}u_{b}.
\end{eqnarray}
 For the metric of the form eq. (\ref{eq3}) and considering a perfect fluid in a flat goemetry, the Ricci scalar and Gauss-Bonnet parameter are presented as
\begin{eqnarray}
 && R=6\Big(2H^{2}+\dot{H}\Big),\\
 && G=24H^{2}\Big(H^{2}+\dot{H}\Big),\label{eq17}
\end{eqnarray}
where $H=\frac{\dot{a}}{a}$ is the Hubble parameter. The dot describes differentiation with respect to cosmic time $t$.  The variation principle in the gravitational action with respect to the metric $g^{\mu\nu}$, the scalar field $\phi$, and $G$  produces the field equations for $f(G)$ gravity which can be presented as
\begin{eqnarray}
 && 3H^{2}=\rho_{m}+\rho_{\phi}+\rho_{G},\label{eq18}\\
 && -3H^{2}-2\dot{H}=p_{m}+p_{\phi}+p_{G}\label{eq19},
 \end{eqnarray}
where  $\rho_{m}$ and $p_{m}$ represent both relativistic matter (photons, neutrons) and non-relativistic matter (baryons, leptons, Cold Dark Matter ), $\rho_{\phi}$ and $p_{\phi}$ represents the matter contribution from the scalar field, $\rho_{G}$ and $p_{G}$  represents the contribution from the Gauss-Bonnet term.
The energy density for scalar field contribution and for Gauss-Bonnet contribution and the their pressures are presented as \cite{copeland2006dynamics,li2007cosmology,venikoudis2022late}
\begin{eqnarray}
&&\rho_{\phi}=\frac{1}{2}\dot{\phi}^{2}+V(\phi),\label{eq20}\\
&& p_{\phi}=\frac{1}{2}\dot{\phi}^{2}-V(\phi),\\
&&\rho_{G}=\frac{Gf'-f}{2}-12\dot{f}'H^{3},\\
&&p_{G}=\frac{f-Gf'}{2}+8\dot{f}'H\dot{H}+4H^{2}\left(\ddot{f}'+2H\dot{f}'\right)\label{eq23}.
\end{eqnarray}
The matter follows the perfect fluid assumption for any cosmological era of interest. The pressure of the perfect fluid is presented as
\begin{eqnarray}
 p_{m}=w\rho_{m},
\end{eqnarray}
 where $m$ specifies relativistic or non-relativistic matter, whereas the pressure of the total fluids is given by
\begin{eqnarray}
p_{t}=w_{t}\rho_{t}\;,
\end{eqnarray}
where $w_{t}=w_{m}+w_{G}+w_{\phi}=\frac{p_{m}+p_{G}+p_{\phi}}{\rho_{m}+\rho_{G}+\rho_{\phi}}$ and $\rho_{t}=\rho_{m}+\rho_{G}+\rho_{\phi}$.
The continuity equations in the context of scalar field assisted $f(G)$ gravity are given by
\begin{eqnarray}
&&\dot{\rho}_{m}+3H\Big(1+3w_{m}\Big)\rho_{m}=0\;,\\
&&\dot{\rho}_{G}+3H\Big(1+3w_{G}\Big)\rho_{G}=0\;,\\
&&\dot{\rho}_{\phi}+3H\Big(1+3w_{\phi}\Big)\rho_{\phi}=0\;,
\end{eqnarray}
with the equation of state parameter for matter fluid, Gauss-Bonnet and scalar field fluids are given respectively as
\begin{eqnarray}
 w_{m}=\frac{p_{m}}{\rho_{m}}\;,
w_{G}=\frac{p_{G}}{\rho_{G}}\;,
w_{\phi}=\frac{p_{\phi}}{\rho_{\phi}}.
\end{eqnarray}
Throughout this work, we assume $w_{m}$ to be a constant  and $w_{G}$ and $w_{\phi}$ dynamically changes.
 \\
The Raychaudhuri
equation $\dot{\theta}=-\frac{1}{3}\theta^{2}-\frac{1}{2}(\rho+3p)+\bigtriangledown_{a}\dot{u}_{a}$  that governs the expansion history of the universe in this case is given by
\begin{eqnarray}
 &&\dot{\theta}=-\frac{1}{3}\theta^{2}-\frac{1}{2}\Big((1+3w_{m})\rho_{m}+2\dot{\phi}-2V(\phi)-Gf'+f-12f''\dot{G}H^{3}-\frac{f''G\dot{G}}{H}\nonumber\\&&-12H^{2}f''\ddot{G}+12H^{2}f'''\dot{G}^{2}\Big)+\bigtriangledown_{a}\dot{u}_{a},\label{eq30}
\end{eqnarray}
with $\dot{u}_{a}$ as the $4$-acceleration in the energy frame of the total fluids  given by
\begin{eqnarray}
&&\dot{u}_{a}=-\frac{\tilde{\bigtriangledown}_{a}p_{t}}{\rho_{t}+p_{t}}=-\frac{1}{(1+w_{t})\rho_{t}}\Big[(1+w_{m})\tilde{\bigtriangledown}_{a}\rho_{m}+\dot{\phi}\tilde{\bigtriangledown}_{a}\dot{\phi}-\tilde{\bigtriangledown}_{a}V(\phi)+\Big(\frac{Gf''}{3H}\nonumber\\
&&+8H^{2}f''\dot{G}\Big)\tilde{\bigtriangledown}_{a}\dot{G}+\Big(\frac{1}{2}-\frac{Gf''}{2}+\frac{f'}{2}+\frac{G\dot{G}f'''+\dot{Gf''}}{3H}+\frac{4H^{2}f''\dddot{G}}{\dot{G}}+4H^{2}\ddot{G}f'''\nonumber\\
&&+4H^{2}\dot{G}^{2}f'''\Big)\tilde{\bigtriangledown}_{a}G+\Big(8f''H\ddot{G}+8f''H\dot{G}^{2}-\frac{Gf''\dot{G}}{3H^{2}}\Big)\tilde{\bigtriangledown}_{a}H\Big]\;.\label{eq31}
\end{eqnarray}
 In the next section, we define the gradient variables which help us to obtain perturbation  equations governing the evolution of the universe.
\section{Fluid description and linear evolution equations}\label{sec4}
\subsection{Definition of vector gradient variables  for the evolution of the universe}
\subsubsection{Matter fluids\\}
Considering the homogeneous and isotropic expanding FRW cosmological background,   spatial gradient variables namely matter energy overdensity and volume expansion of the universe can be defined as
\begin{eqnarray}
 &&D^{m}_{a}=\frac{a \tilde{\bigtriangledown}_{a}\rho_{m}}{\rho_{m}},\label{eq32}\\
 && Z_{a}=a\tilde{\bigtriangledown}_{a}\theta,\label{eq33}
 \end{eqnarray}
 with $m$ as the running index to specify matter.  \subsubsection{Gauss-Bonnet fluids\\}
 Analogous to the $1+3$ covariant perturbations, the definition of key gradient variables resulting from the spatial gradient variables connected to the Gauss-Bonnet fluids in $f(G)$ gravity can be done as
 \begin{eqnarray}
  && \mathcal{G}=a \tilde{\bigtriangledown}_{a}G,\label{eq34}\\
 && \mathbf{G}=a \tilde{\bigtriangledown}_{a}\dot{G},\label{eq35}
 \end{eqnarray}
 where $\mathcal{G}$ and $\mathbf{G}$ characterize the perturbations due to Gauss-Bonnet invariant $G$ and its momentum $\dot{G}$ and describe the inhomogeneity in the Gauss-Bonnet fluids.
 \subsubsection{Scalar field fluids\\}
 The gradient variables responsible to the perturbations due to scalar field $\phi$ and its momentum $\dot{\phi}$ are presented as
 \begin{eqnarray}
 && \Phi_{a}=a\tilde{\bigtriangledown}_{a}\phi,\label{eq36}\\
 &&\Psi_{a}=a\tilde{\bigtriangledown}_{a}\dot{\phi}\label{eq37}.
\end{eqnarray}
 These defined gradient variables eq. (\ref{eq32}) through to eq. (\ref{eq37}) can be used to develop the system of cosmological perturbations of the scalar field assisted $f(G)$ gravity in the $1+3$ covariant formalism. Using scalar decomposition method, the scalar gradient variables  can be extracted from eq. (\ref{eq32}) through to eq. (\ref{eq37}) and presented as
 \begin{eqnarray}
\Delta_{m}=a\tilde{\bigtriangledown}_{a} D^{m}_{a},
 Z=a\tilde{\bigtriangledown}_{a}Z_{a},
 \mathcal{G}=a\tilde{\bigtriangledown}_{a}\mathcal{G}_{a},
 \mathbf{G}=a\tilde{\bigtriangledown}_{a}\mathbf{G}_{a},\Phi=a\tilde{\bigtriangledown}_{a}\Phi_{a},\Psi =a\tilde{\bigtriangledown}_{a}\Psi_{a}. \label{eq38}
 \end{eqnarray}
The time derivative of each gradient variable helps to know how each evolves and help us to get linear perturbation equations responsible for large scale structure formation in the context of the scalar field assisted $f(G)$ gravity.
\subsection{Derivation of linear perturbation equations}
The time derivative of eq. (\ref{eq38}) and making use of eq. (\ref{eq30}), eq. (\ref{eq31}) and eq. (\ref{a0}) yields
\begin{eqnarray}
 &&\dot{\Delta} _{m}=\frac{(1+w_{m})\theta}{(1+w_{t})\rho_{t}}\Big((1+w_{m})\rho_{m}\Big)\Delta_{m}+b_{13}(1+w_{m})\theta\Psi-b_{14}(1+w_{m})\theta\Phi\nonumber\\&&+\Big[\chi-(1+w_{m})\Big]Z +\mu\mathcal{G}+\nu\mathbf{G}\;, \label{eq39}\\
&&\dot{Z}=\Big(-\frac{2\theta}{3}-\tau\Big)Z+\Big[-\frac{(1+3w_{m})\rho_{m}}{2}-\Upsilon\Big]\Delta_{m}+\Big(2-\gamma\Big)\Psi+\zeta\Phi\nonumber\\&&-\Big(1+\varepsilon \Big)\mathcal{G}-\eta\mathbf{G}-\frac{(1+w_{m})\rho_{m}}{(1+w_{t})\rho_{t}}\bigtriangledown^{2}\Delta_{m}-b_{13}\bigtriangledown^{2}\Psi+b_{14}\bigtriangledown^{2}\Phi-b_{1}\bigtriangledown^{2}\mathbf{G}\nonumber\\&&-b_{6}\bigtriangledown^{2}\mathcal{G}-b_{7}\bigtriangledown^{2}Z,\label{eq40}\\
 &&\dot{\mathcal{G}}=\Big(1-b_{1}\dot{G}\Big)\mathbf{G}-\frac{(1+w_{m})\rho_{m}}{(1+w_{t})\rho_{t}}\dot{G}\Delta_{m}-b_{13}\dot{G}\Psi+b_{14}\dot{G}\Phi-b_{6}\dot{G}\mathcal{G}\nonumber\\&&-b_{7}\dot{G}Z,\label{eq41}\\
  &&\dot{\mathbf{G}}=-b_{2}\mathbf{G}-\frac{(1+w_{m})\rho_{m}}{(1+w_{t})\rho_{t}}\ddot{G}\Delta_{m}-b_{13}\ddot{G}\Psi+b_{14}\ddot{G}\Phi+\iota\mathcal{G}-\xi Z,\label{eq42}\\
 &&\dot{\Phi}=-b_{9}\mathbf{G}-b_{13}(1+w_{m})\rho_{m}\Delta_{m}+b_{12}\Psi+b_{14}\dot{\phi}\Phi-b_{10}\mathcal{G}-b_{11}Z,\label{eq43}\\
 &&\dot{\Psi}=-b_{15}\mathbf{G}-\frac{(1+w_{m})\rho_{m}}{(1+w_{t})\rho_{t}}\ddot{\phi}\Delta_{m}-b_{13}\ddot{\phi}\Psi+b_{16}\ddot{\phi}\Phi-b_{17}\mathcal{G}-b_{18}Z,\label{eq44}
 \end{eqnarray}
 \begin{eqnarray}
 &&\ddot{\Delta} _{m}=\Big[\frac{(1+w_{m})\theta}{(1+w_{t})\rho_{t}}\Big((1+w_{m})\rho_{m}\Big)+\Big[b_{7}b_{13}(1+w_{m})\theta\ddot{\phi}-\Big(\chi\nonumber\\&&-(1+w_{m})\Big)\Big(\frac{2\theta}{3}-\tau\Big)-\mu \xi \Big]\frac{1}{1+w_{m}}\Big]\dot{\Delta}_{m}+\Big[-b_{3}-\frac{(1+3w_{m})\rho_{m}}{2}+\Upsilon\nonumber\\&& -\mu\frac{(1+w_{m})\rho_{m}}{(1+w_{t})\rho_{t}}\ddot{G}-\Big[b_{1}b_{13}(1+w_{m})\theta\ddot{\phi}+\varepsilon+\mu b_{2}\Big]\Big(\frac{w_{m}\dot{G}}{1+w_{m}}\Big)+\Big(-b_{4}\nonumber\\&&-b_{13}\mu\ddot{G}\Big)\Big(\frac{w_{m}\dot{\phi}}{1+w_{m}}\Big)-\Big[b_{7}b_{13}(1+w_{m})\theta\ddot{\phi}-\Big(\chi-(1+w_{m})\Big)\Big(\frac{2\theta}{3}-\tau\Big)\nonumber\\&&-\mu \xi \Big]\Big(\frac{w_{m}\theta}{1+w_{m}}\Big)\Big]\Delta_{m}-\Big[b_{1}b_{13}(1+w_{m})\theta\ddot{\phi}+\varepsilon+\mu b_{2}-\eta\Big]\dot{\mathcal{G}}-\Big[b_{6}b_{13}(1+w_{m})\theta\ddot{\phi}\nonumber\\&&+1+\zeta-\mu \iota\Big]\mathcal{G}-\Big[b_{4}+b_{13}\mu\ddot{G}+b_{14}(1+w_{m})\theta\Big]\dot{\Phi}+\Big(b_{5}+b_{14}\mu\ddot{G}\gamma\Big)\Phi \nonumber\\&&-\Big[\frac{(1+w_{m})\rho_{m}}{(1+w_{t})\rho_{t}}+b_{13}\Big(\frac{w_{m}\dot{\phi}}{1+w_{m}}\Big)-b_{1}\Big(\frac{w_{m}\dot{G}}{1+w_{m}}\Big)+b_{7}\Big(\frac{w_{m}\theta}{1+w_{m}}\Big)\Big]\bigtriangledown^{2}\Delta_{m}\nonumber\\&&-b_{13}\bigtriangledown^{2}\dot{\Phi}+b_{14}\bigtriangledown^{2}\Phi-b_{1}\bigtriangledown^{2}\dot{\mathcal{G}}-b_{6}\bigtriangledown^{2}\mathcal{G}+\frac{b_{7}}{1+w_{m}}\bigtriangledown^{2}\dot{\Delta}_{m},\label{eq45}\\
 &&\ddot{\mathcal{G}}=\Big[\xi\eta-\Big(1-b_{2}\Big)b_{2}+b_{1}b_{13}\dot{G}\ddot{\phi}-\dot{G}b_{6}\Big]\dot{\mathcal{G}}-\{\Big(1-b_{2}\Big)\frac{(1+w_{m})\rho_{m}}{(1+w_{t})\rho_{t}}\ddot{G}\nonumber\\&&+\Big[\xi\eta-\Big(1-b_{2}\Big)b_{2}+b_{1}b_{13}\dot{G}\ddot{\phi}\Big]\frac{w_{m}\dot{G}}{1+w_{m}}+\xi\Big[-\frac{(1+3w_{m})\rho_{m}}{2}+\Upsilon\Big]\nonumber\\&&+b_{13}\dot{G}\frac{(1+w_{m})\rho_{m}}{(1+w_{t})\rho_{t}}\ddot{\phi}+\Big[\Big(1-b_{2}\Big)b_{13}\ddot{G}+\xi\Big(2+\gamma\Big)-b^{2}_{13}\dot{G}\ddot{\phi}\Big]\Big(\frac{w_{m}\dot{\phi}}{1+w_{m}}\Big)\nonumber\\&&+\Big[\Big(1-b_{2}\Big)\xi+\xi\Big(-\frac{2\theta}{3}+\tau\Big)-b_{7}b_{13}\dot{G}\ddot{\phi} \Big]\Big(\frac{w_{m}\theta}{1+w_{m}}\Big)\}\Delta_{m}+\Big[\Big(1-b_{2}\Big)b_{14}\ddot{G}\nonumber\\&&-\xi\zeta-b_{8}\Big]\Phi-\Big[\Big(1-b_{2}\Big)b_{13}\ddot{G}-\xi\Big(2+\gamma\Big)-b^{2}_{13}\dot{G}\ddot{\phi}-b_{14}\dot{G}\Big]\dot{\Phi}\nonumber\\&&+\Big[\Big(1-b_{2}\Big)\iota+\xi\Big(1+\varepsilon\Big)+\xi b_{6}+b_{6}b_{13}\dot{G}\ddot{\phi} \Big]\mathcal{G}+\Big[\Big[\Big(1-b_{2}\Big)\xi+\xi\Big(-\frac{2\theta}{3}+\tau\Big)\nonumber\\&&-b_{7}b_{13}\dot{G}\ddot{\phi} \Big]\Big(\frac{1}{1+w_{m}}\Big)-\frac{(1+w_{m})\rho_{m}}{(1+w_{t})\rho_{t}}\dot{G}\Big]\dot{\Delta}_{m} +\Big[\xi\frac{(1+w_{m})\rho_{m}}{(1+w_{t})\rho_{t}}+b_{13}\xi\frac{w_{m}\dot{\phi}}{1+w_{m}}\nonumber\\&&-\xi b_{1}\frac{w_{m}\dot{G}}{1+w_{m}}+\xi b_{7}\frac{w_{m}\theta}{1+w_{m}}\Big]\bigtriangledown^{2}\Delta_{m}+b_{13}\xi\bigtriangledown^{2}\dot{\Phi}-b_{14}\xi\bigtriangledown^{2}\Phi+\xi b_{1}\bigtriangledown^{2}\dot{\mathcal{G}}\nonumber\\&&-\frac{\xi b_{7}}{1+w_{m}}\bigtriangledown^{2}\dot{\Delta}_{m},\label{eq46}\\
 &&\ddot{\Phi}=\Big[-\Big(b_{2}b_{9}-b_{11}\eta+b_{12}b_{15}\Big)\Big(\frac{w_{m}\dot{G}}{1+w_{m}}\Big)+b_{9}\frac{(1+w_{m})\rho_{m}}{(1+w_{t})\rho_{t}}\ddot{G}+b_{11}\Big(\Upsilon\nonumber\\&&-\frac{(1+3w_{m})\rho_{m}}{2}\Big)-b_{12}\frac{(1+w_{m})\rho_{m}}{(1+w_{t})\rho_{t}}\ddot{\phi}+\Big(b_{9}b_{13}\ddot{G}+b_{11}\Big(2+\gamma\Big)-\nonumber\\&&b_{12}b_{13}\ddot{\phi}\Big)\Big(\frac{w_{m}\dot{\phi}}{1+w_{m}}\Big)-\Big[b_{11}\Big(-\frac{2\theta}{3}+\tau\Big)-b_{9}\xi-b_{12}b_{18}\Big]\Big(\frac{w_{m}\theta}{1+w_{m}}\Big)\Big]\Delta_{m}\nonumber\\&&+\Big(b_{11}\zeta-b_{9}b_{14}\ddot{G}+b_{12}b_{16}\Big)\Phi-\Big[b_{9}\iota+b_{11}\Big(1+\varepsilon\Big)+b_{10}+b_{12}b_{17}\Big] \mathcal{G}\nonumber\\&&+\Big(b_{2}b_{9}-b_{11}\eta+b_{12}b_{15}\Big)\dot{\mathcal{G}}+\Big[b_{14}\dot{\phi}+\Big(b_{9}b_{13}\ddot{G}+b_{11}\Big(2+\gamma\Big)-b_{12}b_{13}\ddot{\phi}\Big)\Big]\dot{\Phi}\nonumber\\&&+\Big[\Big(b_{11}\Big(-\frac{2\theta}{3}+\tau\Big)-b_{9}\xi-b_{12}b_{18}\Big)\Big(\frac{1}{1+w_{m}}\Big)-(1+w_{m})\rho_{m}b_{13}\Big]\dot{\Delta}_{m} \nonumber
 \end{eqnarray}
 \begin{eqnarray}
 &&+\Big[-b_{11}\frac{(1+w_{m})\rho_{m}}{(1+w_{t})\rho_{t}}-b_{11}b_{13}\Big(\frac{w_{m}\dot{\phi}}{1+w_{m}}\Big)-b_{7}b_{11}\Big(\frac{w_{m}\theta}{1+w_{m}}\Big)\nonumber\\&&+b_{1}b_{11}\Big(\frac{w_{m}\dot{G}}{1+w_{m}}\Big)\Big]\bigtriangledown^{2}\Delta_{m}-b_{11}b_{13}\bigtriangledown^{2}\dot{\Phi}+b_{11}b_{14}\bigtriangledown^{2}\Phi\nonumber\\&&-b_{1}b_{11}\bigtriangledown^{2}\dot{\mathcal{G}}-b_{6}b_{11}\bigtriangledown^{2}\mathcal{G}+\frac{b_{7}b_{11}}{1+w_{m}}\bigtriangledown^{2}\dot{\Delta}_{m}\;\label{eq47},
\end{eqnarray}
 where the parameters such as $b_{1}$, $\tau$ are presented in the appendix.
We consider the scalar potential  $V(\phi)$ of the exponential form  \cite{copeland2006dynamics,boehmer2008dynamics}
\begin{eqnarray}
 V(\phi)=V_{0}e^{-\alpha \phi},
\end{eqnarray}
 where $V_{0}$ is the constant, $\phi$ is the scalar field given by \cite{copeland2006dynamics}\footnote{This scalar field gives rise to  the potential giving the power-law expansion,  which may be
used for dark energy and
  possesses cosmological scaling solutions \cite{copeland2006dynamics}},
 \begin{eqnarray}
  \phi=\ln(t),\label{eq49}
 \end{eqnarray}
 and $\alpha=\sqrt{\frac{16\pi}{m}}\frac{1}{M_{p}}$, where $M_{p}$ is the Planck mass, $m$ giving rise to the power-law expansion of the scale factor $a(t)$, so that
\begin{eqnarray}
 a\tilde{\bigtriangledown}_{a}V(\phi)=-\alpha V(\phi)\Phi_{a}\;.
\end{eqnarray} This potential can represent the late-time acceleration phase of the Universe.
For analysis purpose, we apply first the harmonic decomposition method \cite{abebe2012covariant,carloni2006gauge,abebe2015breaking,sahlu2020scalar,munyeshyaka2023multifluid,twagirayezu2023chaplygin}
\begin{eqnarray}
 &&X=\sum_{k}X^{k}(t)Q^{k}(\vec{x}),\\
 &&Y=\sum_{k}Y^{k}(t)Q^{k}(\vec{x}),\\
 &&\tilde{\bigtriangledown}^{2}Q(x)=-\frac{k^{2}}{a^{2}}Q^{k}(x)\;,
\end{eqnarray}
where $k$ is the wave number and $Q^{k}(\vec{x})$ is the eigen functions of the covariant derivative. The wave number $k$ specifies the order of harmonic oscillator related with cosmological scale factor as $k=\frac{2\pi a}{\lambda}$, where $\lambda$ is the physical wavelength of perturbation.
Then we  appy the redshift transformation method using  \cite{sahlu2020scalar,venikoudis2022late,hough2021confronting,sahlu2020perturbations,munyeshyaka2021cosmological,munyeshyaka20231+,munyeshyaka2023perturbations}
\begin{eqnarray}
&& a=\frac{1}{1+z},\\
&& \dot{f}=-(1+z)Hf',\\
&& \ddot{f}=(1+z)^{2}H\Big(\frac{dH}{dz}\frac{df}{dz}+H\frac{d^{2}f}{dz^{2}}\Big)+(1+z)H^{2}\frac{df}{dz},
\end{eqnarray}
so that eq. (\ref{eq39})-(\ref{eq47}) with little algebra can be represented in redshift space as
 \begin{eqnarray}
 &&(1+z)^{2}H^{2}\Delta''_{m}=-\Big[\Big((1+z)H'+H\Big)+\frac{(1+w_{m})\theta}{(1+w_{t})\rho_{t}}\Big((1+w_{m})\rho_{m}\Big)+\Big[b_{7}b_{13}(1+w_{m})\theta l_{2}\nonumber\\ &&+\Big(\chi-(1+w_{m})\Big)\Big(-\frac{2\theta}{3}+\tau\Big)-\mu \xi \Big]\frac{1}{1+w_{m}}+\frac{b_{7}}{1+w_{m}}\frac{k^{2}}{a^{2}}\Big](1+z)H\Delta'_{m}+\Big[-b_{3}\nonumber\\ &&-\frac{(1+3w_{m})\rho_{m}}{2}+\Upsilon -\mu\frac{(1+w_{m})\rho_{m}}{(1+w_{t})\rho_{t}}l_{4}-\Big[b_{1}b_{13}(1+w_{m})\theta l_{2}+\varepsilon+\mu b_{2}\Big]\Big(\frac{w_{m}l_{3}}{1+w_{m}}\Big)\nonumber\\ &&-\Big(b_{4}+b_{13}\mu l_{4}\Big)\Big(\frac{w_{m}l_{2}}{1+w_{m}}\Big)-\Big[b_{7}b_{13}(1+w_{m})\theta l_{2}+\Big(\chi-(1+w_{m})\Big)\Big(-\frac{2\theta}{3}+\tau\Big)\nonumber\\ &&-\mu \xi \Big]\Big(\frac{w_{m}\theta}{1+w_{m}}\Big)-\Big[\frac{(1+w_{m})\rho_{m}}{(1+w_{t})\rho_{t}}+b_{13}\Big(\frac{w_{m}l_{1}}{1+w_{m}}\Big)-b_{1}\Big(\frac{w_{m}l_{3}}{1+w_{m}}\Big)+b_{7}\Big(\frac{w_{m}\theta}{1+w_{m}}\Big)\Big]\frac{k^{2}}{a^{2}}\Big]\Delta^{k}_{m}\nonumber\\ &&+\Big[b_{1}b_{13}(1+w_{m})\theta l_{2}+\varepsilon+\mu b_{2}-\eta-b_{1}\frac{k^{2}}{a^{2}}\Big](1+z)H\mathcal{G}'-\Big[b_{6}b_{13}(1+w_{m})\theta l_{2}+1+\zeta-\mu \iota\nonumber\\ &&-b_{6}\frac{k^{2}}{a^{2}}\Big]\mathcal{G}^{k}+\Big[b_{4}+b_{13}\mu l_{4}+b_{14}(1+w_{m})\theta -b_{13}\frac{k^{2}}{a^{2}}\Big](1+z)H\Phi'+\nonumber\\&&\Big(b_{5}+b_{14}\mu l_{4}\gamma+b_{14}\frac{k^{2}}{a^{2}}\Big)\Phi^{k}\;\label{eq59},\\
   &&(1+z)^{2}H^{2}\mathcal{G}''=-\Big[\Big((1+z)H'+H\Big)+\xi\eta-\Big(1-b_{2}\Big)b_{2}+b_{1}b_{13}l_{3}l_{2}-l_{3}b_{6}\nonumber\\ &&+\xi b_{1}\frac{k^{2}}{a^{2}}\Big](1+z)H\mathcal{G}'-\{\Big(1-b_{2}\Big)\frac{(1+w_{m})\rho_{m}}{(1+w_{t})\rho_{t}} l_{4}+\Big[\xi\eta-\Big(1-b_{2}\Big)b_{2}+b_{1}b_{13}l_{3}l_{2}\Big]\frac{w_{m}l_{3}}{1+w_{m}}\nonumber\\ &&-\xi\Big[\frac{(1+3w_{m})\rho_{m}}{2}-\Upsilon \Big]+b_{13}l_{3}\frac{(1+w_{m})\rho_{m}}{(1+w_{t})\rho_{t}}l_{2}+\Big[\Big(1-b_{2}\Big)b_{13}l_{4}+\xi\Big(2+\gamma\Big)\nonumber\\ &&-b^{2}_{13}l_{3}l_{2}\Big]\Big(\frac{w_{m}l_{1}}{1+w_{m}}\Big)+\Big[\Big(1-b_{2}\Big)\xi+\xi\Big(-\frac{2\theta}{3}+\tau \Big)-b_{7}b_{13}l_{3}l_{2} \Big]\Big(\frac{w_{m}\theta}{1+w_{m}}\Big)\nonumber\\ &&+\Big[\xi \frac{(1+w_{m})\rho_{m}}{(1+w_{t})\rho_{t}}+b_{13}\xi\frac{w_{m}l_{1}}{1+w_{m}}-\xi b_{1}\frac{w_{m}l_{3}}{1+w_{m}}+\xi b_{7}\frac{w_{m}\theta}{1+w_{m}}\Big]\frac{k^{2}}{a^{2}}\}\Delta^{k}_{m}+\Big[\Big(1-b_{2}\Big)b_{14}l_{4}\nonumber\\ &&-\xi\zeta-b_{8}-b_{14}\xi\frac{k^{2}}{a^{2}} \Big]\Phi^{k}+\Big[\Big(1-b_{2}\Big)b_{13}l_{4}-\xi\Big(2+\gamma\Big)-b^{2}_{13}l_{3}l_{2}-b_{14}l_{3}+b_{13}\xi\frac{k^{2}}{a^{2}}\Big](1+z)H\Phi'\nonumber\\&&+\Big[\Big(1-b_{2}\Big)\iota+\xi\Big(1+\varepsilon\Big)+\xi b_{6}+b_{6}b_{13}l_{3}l_{2} \Big]\mathcal{G}^{k}-\Big[\Big[\Big(1-b_{2}\Big)\xi+\xi\Big(-\frac{2\theta}{3}+\tau\Big)\nonumber\\&&-b_{7}b_{13}l_{3}l_{2} \Big]\Big(\frac{1}{1+w_{m}}\Big)-\frac{(1+w_{m})\rho_{m}}{(1+w_{t})\rho_{t}}l_{3}-\frac{\xi b_{7}}{1+w_{m}}\frac{k^{2}}{a^{2}}\Big](1+z)H\Delta'_{m}\;,\label{eq60}
   \end{eqnarray} and
 \begin{eqnarray}
 &&(1+z)^{2}H^{2}\Phi''=\Big[-\Big(b_{2}b_{9}-b_{11}\eta+b_{12}b_{15}\Big)\Big(\frac{w_{m}l_{3}}{1+w_{m}}\Big)+b_{9}\frac{(1+w_{m})\rho_{m}}{(1+w_{t})\rho_{t}}l_{4}+b_{11}\Big(\Upsilon\nonumber\\ &&-\frac{(1+3w_{m})\rho_{m}}{2}\Big)-b_{12}\frac{(1+w_{m})\rho_{m}}{(1+w_{t})\rho_{t}}l_{2}+\Big(b_{9}b_{13}l_{4}+b_{11}\Big(2+\gamma\Big)-b_{12}b_{13}l_{2}\Big)\Big(+\frac{w_{m}l_{1}}{1+w_{m}}\Big)\nonumber\\&&-\Big[b_{11}\Big(-\frac{2\theta}{3}+\tau\Big)-b_{9}\xi-b_{12}b_{18}\Big]\Big(\frac{w_{m}\theta}{1+w_{m}}\Big)+\Big[-b_{11}\frac{(1+w_{m})\rho_{m}}{(1+w_{t})\rho_{t}}-b_{11}b_{13}\Big(\frac{w_{m}l_{1}}{1+w_{m}}\Big)\nonumber\\&&-b_{7}b_{11}\Big(\frac{w_{m}\theta}{1+w_{m}}\Big)+b_{1}b_{11}\Big(\frac{w_{m}l_{3}}{1+w_{m}}\Big)\Big]\frac{k^{2}}{a^{2}}\Big]\Delta^{k}_{m}+\Big(b_{11}\zeta-b_{9}b_{14}l_{3}+b_{12}b_{16}+b_{11}b_{14}\frac{k^{2}}{a^{2}}\Big)\Phi^{k}\nonumber\\&&-\Big[b_{9}\iota+b_{11}\Big(1+\varepsilon\Big)+b_{10}+b_{12}b_{17}-b_{6}b_{11}\frac{k^{2}}{a^{2}}\Big] \mathcal{G}^{k}-\Big(b_{2}b_{9}-b_{11}\eta+b_{12}b_{15}-b_{1}b_{11}\frac{k^{2}}{a^{2}}\Big)(1+z)H\mathcal{G}'\nonumber\\&&-\Big[\Big((1+z)H'+H\Big)+b_{14}l_{1}+\Big(b_{9}b_{13}l_{4}+b_{11}\Big(2+\gamma\Big)-b_{12}b_{13}l_{4}\Big)-b_{11}b_{13}\frac{k^{2}}{a^{2}}\Big](1+z)H\Phi'\nonumber\\&&-\Big[\Big(b_{11}\Big(-\frac{2\theta}{3}+\tau\Big)-b_{9}\xi-b_{12}b_{18}\Big)\Big(\frac{1}{1+w_{m}}\Big)-(1+w_{m})\rho_{m}b_{13}+\frac{b_{7}b_{11}}{1+w_{m}}\frac{k^{2}}{a^{2}}\Big](1+z)H\Delta'_{m}\;. \label{eq61}
\end{eqnarray}
The   Eq. (\ref{eq59}) through to Eq. (\ref{eq61}) remain key equations for analysing the growth of energy density fluctuations for the purpose of explaining the  large scale structures formation.  For pedagogical purpose and for the sake of simplicity, we consider three different  $f(G)$ models, namely model$1$, model$2$ and model$3$  \footnote{ These models were  choosen for a quantitative
analysis of the evolution of cosmological perturbations in $f (G)$ gravity.
they are  viable models that may
represent examples of models that could account for the
late-time acceleration of the universe without the need for dark energy \cite{venikoudis2022late}},
\begin{eqnarray}
 &&f(G)=\frac{\alpha_{1}}{G^{2}}+\alpha_{2}G^{\frac{1}{2}},\nonumber\\
 &&f(G)=f_{0}\textbardbl{G}^{\beta}\nonumber\\
 &&f(G)=\frac{c_{1}G^{n}+d_{1}}{c_{2}G^{n}+d_{2}},\label{eq62}
\end{eqnarray}  respectively considered in \cite{venikoudis2022late}, \cite{nojiri2007dark,nojiri2011unified,nojiri2005modified} and \cite{nojiri2008inflation,bamba2010finite},
where $\alpha_{1}$, $\alpha_{2}$, $\beta$, $f_{0}$, $c_{1}$
$c_{2}$, $d_{1}$ and $d_{2}$ are constants and $\beta>0$, $n>0$ and $G\neq0$,   and the scalar field (eq. \ref{eq49}) with its derivatives ($l_{1}$, $l_{2}$, $l_{6}$), the Hubble parameter  and the  Gauss-Bonnet parameter (eq. \ref{eq17}) with its derivatives ($l_{3}$, $l_{4}$, $l_{5}$) presented in redshift space as
\begin{eqnarray}
 &&
\phi(z)=\frac{2m}{3(1+w)}\phi \ln(1+z),\label{eq63}\\
&&H=\frac{2m}{3(1+w)}(1+z)^{\frac{3(1+w)}{2m}},\label{eq64}\\
&&G=\beta_{1}(1+z)^{\frac{6(1+w)}{m}}\label{eq65},
\end{eqnarray}
 with $\beta_{1}=24(\frac{2m}{3(1+w)})^{4}\Big(1-\frac{3(1+w)}{2m}\Big)$, $l_{1}=-\frac{2m}{3(1+w)}H$, $l_{2}=\frac{2m}{3(1+w)}(1+z)HH'$, $l_{3}=-4\beta_{1}(1+z)^{\frac{15(1+w)}{2m}}$, $l_{4}=-20\beta_{1}(1+z)^{\frac{9(1+w)}{m}}$, $l_{5}=80\beta_{1}(1+z)^{\frac{21(1+w)}{2m}}$ and $l_{6}=-H$ to analyse the growth of matter fluctuations in the context of the scalar field assisted $f(G)$ gravity and GR limits. For different realistic $f(G)$ models, the work presented in \cite{bamba2008future,bamba2010finite,de2009construction,bamba2017energy} are recommended.
\section{Matter density fluctuations in GR}\label{sec5}
For the case $f(G)=G$, ( when $\alpha_{1}\sim 0$ and $\alpha_{2}\sim G^{\frac{1}{2}}$, $f_{0}\sim 1$ and $\beta \sim 1$, $c_{1}=d_{2}=1$, $n=1$ and $d_{1}=c_{2}=0$, from eq. (\ref{eq62})) and without considering the scalar field contribution in  Eq. (\ref{eq59}) through to Eq. (\ref{eq61}), we obtain
\begin{eqnarray}
 &&\Delta''_{m}+\Big[H'+\frac{4+3w_{m}}{(1+z)H}+\frac{2}{1+z}\Big]\Delta'_{m}+\Big[\frac{6w_{m}}{(1+z)^{2}H}-\frac{1+3w_{m}}{2(1+z)}-\frac{k^{2}}{((1+z)aH)^{2}}\Big]\Delta^{k}_{m}=0,
 \label{eq66}\\
 &&\mathcal{G}''=0,\\
 &&\Phi''=0,
 \label{eq68}
\end{eqnarray} which coincides with GR limits. In this case, the matter (relativistic and non-relativistic) is involved in the growth of the energy density fluctuations. The evolution of Eq. (\ref{eq59}) through to Eq. (\ref{eq61})
 coincides with the GR limits. We analyse the growth of energy density fluctuations for both dust and radiation fluids in GR limits (eq. \ref{eq66}--\ref{eq68}).
 \subsection{Dust-dominated epoch}
\begin{figure}
  \includegraphics[width=100mm]{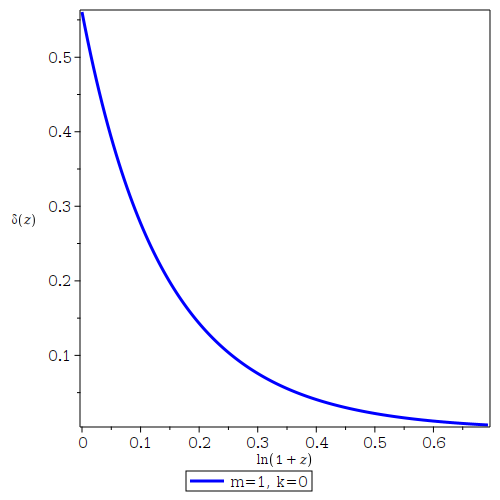}
  \caption{Plot of energy density perturbations versus redshift of Eq. (\ref{eq69}) for dust-dominated universe.}
  \label{Fig1}
  \end{figure}

 We assume the universe is only dominated by dust fluid, with equation of state parameter $w=0$, therefore eq. (\ref{eq66})-(\ref{eq68}) lead to
 \begin{eqnarray}
 \Delta''_{d}+\Big[H'_{d}+\frac{4}{(1+z)H_{d}}+\frac{2}{1+z}\Big]\Delta'_{d}-\Big[\frac{1}{2(1+z)}+\frac{k^{2}}{((1+z)aH_{d})^{2}}\Big]\Delta_{d}=0,
 \label{eq69}
\end{eqnarray}
 where the subscript $d$ stands for dust. Eq. (\ref{eq69})  admits the numerical results which  are presented in fig. (\ref{Fig1}),
  where we depict that the matter energy density fluctuations decay with increase in redshift.
  \subsection{Radiation-dominated epoch}
\begin{figure}
  \includegraphics[width=100mm]{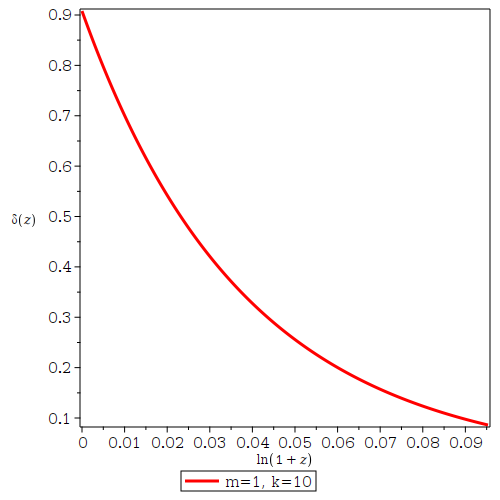}
  \caption{Plot of energy density perturbations versus redshift of eq. (\ref{eq70})  for radiation-dominated universe.}
  \label{Fig2}
  \end{figure}

  By assuming a universe dominated by radiation fluid, with equation of state parameter $w=\frac{1}{3}$,  eq. (\ref{eq66}--\ref{eq68}) lead to
\begin{eqnarray}
 &&\Delta''_{r}+\Big[H'_{r}+\frac{5}{(1+z)H_{r}}+\frac{2}{1+z}\Big]\Delta'_{r}-\Big[\frac{1}{(1+z)}-\frac{2}{(1+z)^{2}H_{r}}+\frac{k^{2}}{((1+z)aH_{r})^{2}}\Big]\Delta^{k}_{r}=0,
 \label{eq70}
\end{eqnarray}
 where $r$ stands for radiation. Eq. (\ref{eq70}) admits the numerical solution presented in fig. (\ref{Fig2}).
  During numerical results computation, the relation $\frac{k^{2}}{(aH)^{2}}=\frac{16\pi^{2}}{\lambda^{2}}{(1+z)^{4}}$ is useful, and the normalised energy density fluctuations can be defined as
  \begin{eqnarray}
   \delta(z)=\frac{\Delta(z)}{\Delta(z_{in})},
  \end{eqnarray}
where $\Delta(z_{in})$ is the initial value of $\Delta(z)$ at $z_{in}=1100$. Since the variation of CMB temperature detected observationally is in the order of $10^{-5}$ at $z_{in}=1100$, we use the initial conditions for $\Delta(z)=10^{-5}$ and $\Delta'(z)=0$.
   From the plots, one can depict that, the matter energy density fluctuations for both dust- and radiation-dominated epochs are decaying with increase in redshift for different values of $k$.
\section{Matter density fluctuations of the scalar field assisted $f(G)$ gravity}\label{sec6}
In this section, we analyze the growth of energy density fluctuations of the perturbation equations in both short- and long-wavelength  modes by considering the case where the universe is dominated by the mixture of dust-scalar field-Gauss-Bonnet fluids and the case where the mixture of radiation-scalar field-Gauss-Bonnet fluids is dominating the universe.
\subsection{Short-wavelength mode}
The growth of fractional energy density perturbations is analyzed within the Hubble horizon, where $\frac{k^{2}}{a^{2}H^{2}}\gg 1$ in the whole system to present numerical results of energy density fluctuations. The work done in \cite{dunsby1991gauge} for GR, \cite{abebe2012covariant} for $f(R)$, \cite{sahlu2020scalar} for $f(T)$ and in \cite{munyeshyaka2021cosmological,twagirayezu2023chaplygin,munyeshyaka2023multifluid} for $f(G)$ considered this limit.
\subsubsection{Quasi-static approximation}
In the quasi-static approximation, the time fluctuations in the perturbations of the scalar-field energy density and its momentum, and that of the Gauss-Bonnet energy density  and its momentum are assumed to be constant with time. In this case, one sets $\dot{\mathcal{G}}=\dot{\mathbf{G}}=\dot{\Phi}=\dot{\Psi}=0$ under this approximation, the linear perturbations equations eq. (\ref{eq59}) through to eq. (\ref{eq61}) reduce to
\begin{eqnarray}
 &&(1+z)^{2}H^{2}\Delta''_{m}=-\Big[\Big[\Big((1+z)H'+H\Big)+\frac{(1+w_{m})\theta}{(1+w_{t})\rho_{t}}\Big((1+w_{m})\rho_{m}\Big)\nonumber\\ &&+\Big[b_{7}b_{13}(1+w_{m})\theta l_{2}+\Big(\chi-(1+w_{m})\Big)\Big(-\frac{2\theta}{3}+\tau\Big)\nonumber\\&&-\mu \xi \Big]\frac{1}{1+w_{m}}+\frac{b_{7}}{1+w_{m}}\frac{k^{2}}{a^{2}}\Big](1+z)H+E_{2}-E_{4}\Big]\Delta'_{m}+\Big[-b_{3}-\frac{(1+3w_{m})\rho_{m}}{2}+\Upsilon \nonumber\\&&-\mu\frac{(1+w_{m})\rho_{m}}{(1+w_{t})\rho_{t}}l_{4}-\Big[b_{1}b_{13}(1+w_{m})\theta l_{2}+\varepsilon+\mu b_{2}\Big]\Big(\frac{w_{m}l_{3}}{1+w_{m}}\Big)-\Big(b_{4}+b_{13}\mu l_{4}\Big)\Big(\frac{w_{m}l_{2}}{1+w_{m}}\Big)\nonumber\\&&-\Big[b_{7}b_{13}(1+w_{m})\theta l_{2}+\Big(\chi-(1+w_{m})\Big)\Big(-\frac{2\theta}{3}+\tau\Big)-\mu \xi \Big]\Big(\frac{w_{m}\theta}{1+w_{m}}\Big)-\Big[\frac{(1+w_{m})\rho_{m}}{(1+w_{t})\rho_{t}}\nonumber\\&&+b_{13}\Big(\frac{w_{m}l_{1}}{1+w_{m}}\Big)-b_{1}\Big(\frac{w_{m}l_{3}}{1+w_{m}}\Big)+b_{7}\Big(\frac{w_{m}\theta}{1+w_{m}}\Big)\Big]\frac{k^{2}}{a^{2}}+E_{1}-E_{3}\Big]\Delta^{k}_{m}\;,\label{eq72}
 \end{eqnarray}
 where the parameters such as $E_{1}$, $l_{1}$ etc are presented in the appendix. In the following subsections, we consider the cases, where cosmic medium is dominated by a mixture of non interacting fluids, namely dust-scalar field-Gauss-Bonnet fluid and that of radiation-scalar field-Gauss-Bonnet fluid to analyse the energy density perturbations within the Hubble horizon.
 \subsubsection{Perturbations in dust-scalar field-Gauss-Bonnet fluids dominated Universe}
\begin{figure}
  \includegraphics[width=120mm]{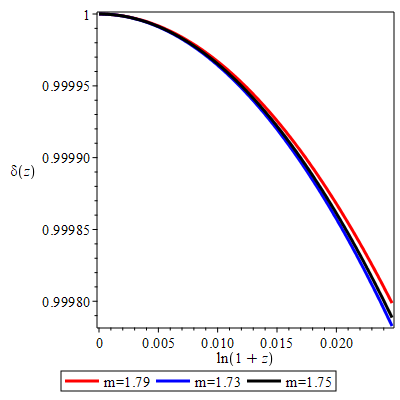}
  \caption{Plot of energy density perturbations versus redshift of Eq. (\ref{eq73}) using model $1$ for different values of $m$ of the dust-scalar field-Gauss-Bonnet fluids system in a short wavelength mode..}
  \label{Fig3}
  \end{figure}

\begin{figure}
  \includegraphics[width=100mm]{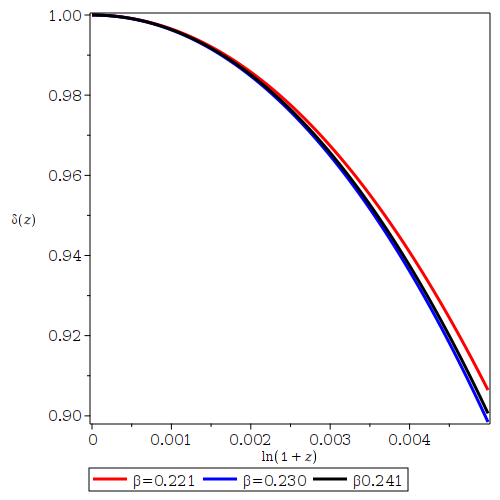}
  \caption{Plot of energy density perturbations versus redshift of Eq. (\ref{eq73}) using model $2$ for different values of $\beta$ of the dust-scalar field-Gauss-Bonnet fluids system in a short wavelength mode..}
  \label{Fig4}
  \end{figure}

\begin{figure}
  \includegraphics[width=100mm]{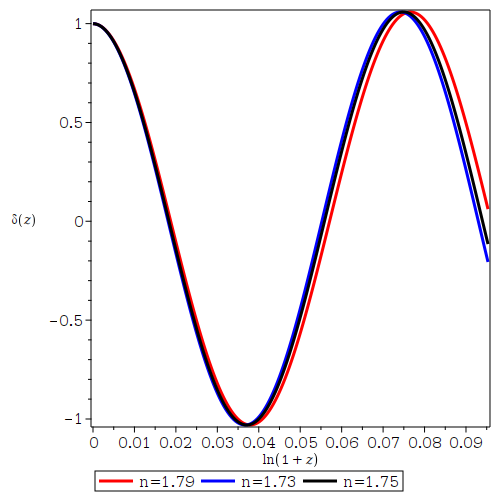}
  \caption{Plot of energy density perturbations versus redshift of Eq. (\ref{eq73}) using model $3$ for different values of $n$ of the dust-scalar field-Gauss-Bonnet fluids system in a short wavelength mode.}
  \label{Fig5}
  \end{figure}

We set the equation of state parameter for dust to be equal to  $w=0$ in the eq. (\ref{eq72}), to get
\begin{eqnarray}
 &&(1+z)^{2}H^{2}\Delta''_{m}=-\Big[(1+z)H'+H+\Big[\frac{\theta \rho_{d}}{(1+w_{t})\rho_{t}}+b_{7}b_{13}\theta l_{2}+\Big(\chi-1\Big)\Big(-\frac{2\theta}{3}+\tau\Big)\nonumber\\&&-\mu \xi +\frac{b_{7}k^{2}}{a^{2}}\Big](1+z)H+E_{4}-E_{2}\Big]\Delta'_{m}+\Big[-b_{3}-\frac{\rho_{d}}{2}+\Upsilon -\mu\frac{\rho_{d}}{(1+w_{t})\rho_{t}}l_{4}\nonumber\\&&-\frac{\rho_{d}}{(1+w_{t})\rho_{t}}\frac{k^{2}}{a^{2}}+E_{1}-E_{3}\Big]\Delta^{k}_{m}\;.\label{eq73}
 \end{eqnarray} The matter energy density perturbations in eq. (\ref{eq73}) do not  couple with the energy density perturbations resulting from the scalar field and the Gauss-Bonnet fluids. Eq. (\ref{eq73})
    admits the numerical solutions presented in figs. (\ref{Fig3}), (\ref{Fig4}) and (\ref{Fig5}) for $f(G)$ model $1$, model $2$ and model $3$, respectively.
During numerical computation, we used eqs. (\ref{eq62}-\ref{eq65}) together with $l_{1}-l_{6}$. We chose the initial conditions $\Delta(z)=10^{-5}$ and $\Delta'(z)=0$. The implications of different models on the energy overdensity contrast are presented in table (\ref{table1}) for dust-scalar field-Gauss-Bonnet fluid mixture.
\begin{table}[ht]
\caption{Illustration of the features of energy overdensity contrast for dust-scalar field-Gauss-Bonnet fluids system in  short wavelength mode.}
\centering
\begin{tabular}{c c c c }
\hline
   Models &Ranges of parameters for  $f(G)$ models  &  behavior of $\delta(z)$ & Observation \\ [0.5ex]

\hline
 model$1$ & $\alpha_{1}=100$, $\alpha_{2}=0.001$ & decay with increase in $z$& observationally supported   \\    \hline

 model$2$ &$0<\beta<\frac{1}{2}$ & decay with increase in  $z$ & observational supported  \\   \hline

 model$3$ &$1<n<2$ & oscillates & observationally supported  \\[1ex]
\hline
\end{tabular}
\label{table1}
\end{table}
\subsubsection{Perturbations in radiation-scalar field-Gauss-Bonnet fluids dominated Universe}
\begin{figure}
  \includegraphics[width=100mm]{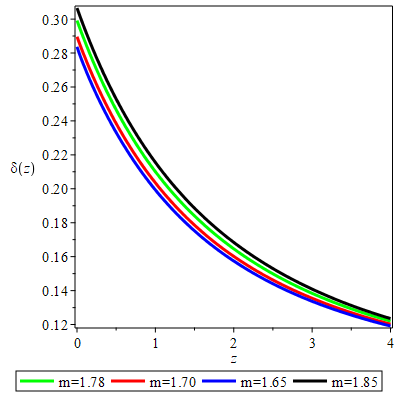}
  \caption{Plot of energy density perturbations versus redshift of Eq. (\ref{eq74}) using model $1$ for different values of $m$ of the radiation-scalar field-Gauss-Bonnet fluids system in a short wavelength mode.}
  \label{Fig6}
  \end{figure}

\begin{figure}
  \includegraphics[width=100mm]{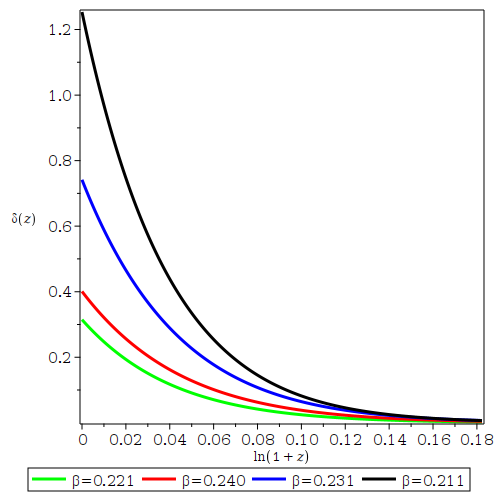}
  \caption{Plot of energy density perturbations versus redshift of Eq. (\ref{eq74}) using model $2$ for different values of $\beta$ of the radiation-scalar field-Gauss-Bonnet fluids system in a short wavelength mode.}
  \label{Fig7}
  \end{figure}

\begin{figure}
  \includegraphics[width=100mm]{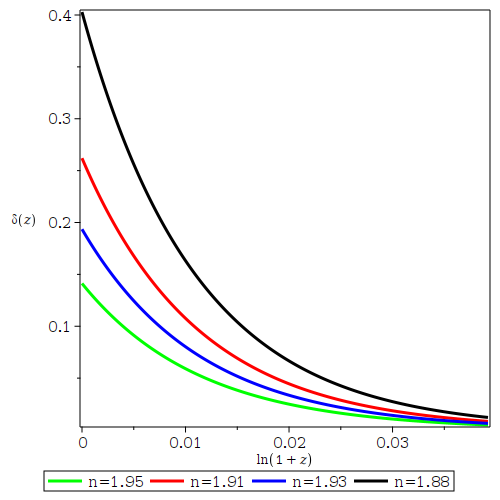}
  \caption{Plot of energy density perturbations versus redshift of Eq. (\ref{eq74}) using model $3$ for different values of $n$ of the radiation-scalar field-Gauss-Bonnet fluids system in a short wavelength mode.}
  \label{Fig8}
  \end{figure}

For the case of radiation-dominated universe, we set the equation of state parameter to $w=\frac{1}{3}$ in eq. (\ref{eq72}) to get
\begin{eqnarray}
 &&(1+z)^{2}H^{2}\Delta''_{m}=-\Big[(1+z)H'+H+\Big[\frac{16\theta \rho_{r}}{9(1+w_{t})\rho_{t}}\nonumber\\ &&+ b_{7}b_{13}\theta l_{2}+\frac{3}{4}\Big(\chi-\frac{4}{3} \Big)\Big(-\frac{2\theta}{3}+\tau\Big)-\frac{3}{4}\mu \xi +\frac{3b_{7}}{4}\frac{k^{2}}{a^{2}}\Big](1+z)H+E_{4}-E_{2}\Big]\Delta'_{m}\nonumber\\&&+\Big[-b_{3}-\rho_{r}+\Upsilon -\mu\frac{4\rho_{r}}{3(1+w_{t})\rho_{t}}l_{4}-\Big[\frac{4}{3} b_{1}b_{13}\theta l_{2}+\varepsilon+\mu b_{2}\Big]\Big(\frac{l_{3}}{4}\Big)-\Big(b_{4}+b_{13}\mu l_{4}\Big)\Big(\frac{l_{2}}{4}\Big)\nonumber\\&&-\Big[\frac{4}{3} b_{7}b_{13}\theta l_{2}+\Big(\chi-\frac{4}{3} \Big)\Big(-\frac{2\theta}{3}+\tau\Big)-\mu \xi \Big]\Big(\frac{\theta}{4}\Big)-\Big[\frac{4\rho_{r}}{3(1+w_{t})\rho_{t}}\nonumber\\&&+b_{13}\Big(\frac{l_{1}}{4}\Big)-b_{1}\Big(\frac{l_{3}}{4}\Big)+b_{7}\Big(\frac{\theta}{4}\Big)\Big]\frac{k^{2}}{a^{2}}+E_{1}-E_{3}\Big]\Delta^{k}_{m}\;.\label{eq74}
 \end{eqnarray} The matter energy density perturbations in eq. (\ref{eq74}) do not  couple with the energy density perturbations resulting from the scalar field and the Gauss-Bonnet fluids. Eq. (\ref{eq74})  admits the numerical solutions presented in figs. (\ref{Fig6}), (\ref{Fig7}), (\ref{Fig8}) for $f(G)$ model $1$, model $2$, model $3$, respectively. The energy density perturbations decay with increase in redshift. The implications of different models on the energy overdensity contrast are presented in table (\ref{table2}) for radiation-scalar field-Gauss-Bonnet fluid mixture in the short wavelength mode.
  \begin{table}[ht]
\caption{Illustration of the features of energy overdensity contrast for radiation-scalar field-Gauss-Bonnet fluids system in  short wavelength mode.}
\centering
\begin{tabular}{c c c c }
\hline
  Models & Ranges of parameters for $f(G)$ models  &  behavior of $\delta(z)$ & Observation \\ [0.5ex]

\hline
 model$1$ &$\alpha_{1}=1$, $\alpha_{2}=1$ & decay with  increase in $z$& observationally supported   \\    \hline

model$2$& $0<\beta<\frac{1}{2}$ & decay with increase in $z$ & observationally supported  \\   \hline

model$3$& $1<n<2$ & decay with increase in $z$ & observationally supported  \\[1ex]
\hline
\end{tabular}
\label{table2}
\end{table}

\subsection{Long-wavelength modes}
 In the long wavelength modes, we assume $\frac{k^{2}}{a^{2}H^{2}}\ll 1$. All cosmological fluctuations begin and remain inside the Hubble horizon. In this case, we solve the whole system of perturbation equations without considering the quasi-static approximation and present the numerical results in both dust-scalar field-Gauss-Bonnet and radiation-scalar field-Gauss-Bonnet fluid mixtures.
\subsubsection{Perturbations in dust-scalar field-Gauss-Bonnet fluids dominated Universe}
\begin{figure}
  \includegraphics[width=100mm]{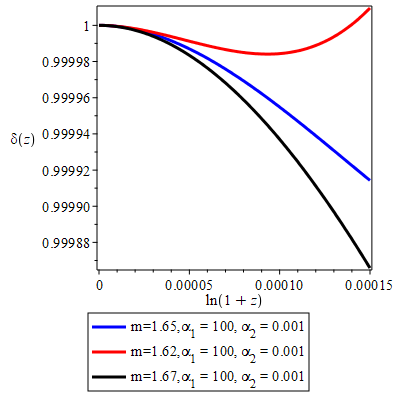}
  \caption{Plot of energy density perturbations versus redshift of Eq. (\ref{eq75}-\ref{eq77}) using model $1$ for different values of $m$ of the dust-scalar field-Gauss-Bonnet fluids system in a long wavelength mode.}
  \label{Fig9}
  \end{figure}

\begin{figure}
  \includegraphics[width=100mm]{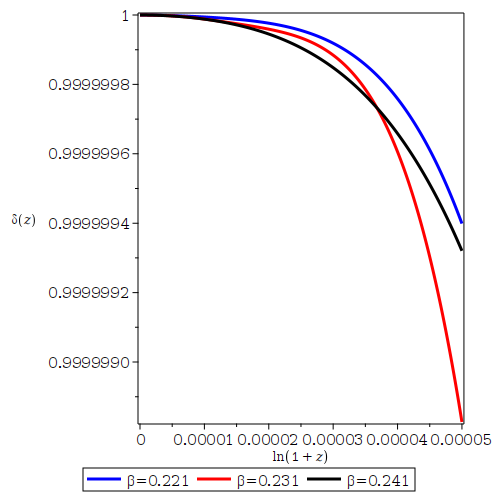}
  \caption{Plot of energy density perturbations versus redshift of Eq. (\ref{eq75}-\ref{eq77}) using model $2$ for different values of $\beta$ of the dust-scalar field-Gauss-Bonnet fluids system in a long wavelength mode.}
  \label{Fig10}
  \end{figure}

\begin{figure}
  \includegraphics[width=100mm]{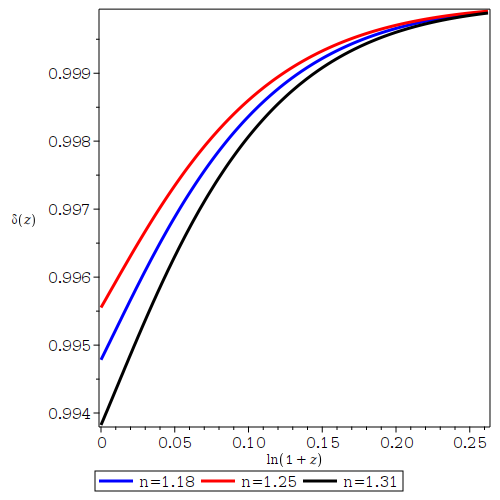}
  \caption{Plot of energy density perturbations versus redshift of Eq. (\ref{eq75}-\ref{eq77}) using model $3$ for different values of $n$ of the dust-scalar field-Gauss-Bonnet fluids system in a long wavelength mode.}
  \label{Fig11}
  \end{figure}

Considering the dust-dominated universe, where $w=0$, eq. (\ref{eq59}) through to eq. (\ref{eq61}) reduce to
\begin{eqnarray}
 &&(1+z)^{2}H^{2}\Delta''_{m}=-\Big[(1+z)H'+H+\frac{\theta}{(1+w_{t})\rho_{t}}\rho_{m}+b_{7}b_{13}\theta l_{2}+\Big(\chi-1\Big)\Big(-\frac{2\theta}{3}+\tau\Big)\nonumber\\ &&-\mu \xi \Big](1+z)H\Delta'_{m}-\Big[b_{3}+\frac{\rho_{m}}{2}-\Upsilon +\mu\frac{\rho_{m}}{(1+w_{t})\rho_{t}}l_{4}\Big]\Delta^{k}_{m}+\Big[b_{1}b_{13}\theta l_{2}+\varepsilon+\mu b_{2}\nonumber\\ &&-\eta\Big](1+z)H\mathcal{G}'-\Big[b_{6}b_{13}\theta l_{2}+1+\zeta-\mu \iota\Big]\mathcal{G}^{k}+\Big[b_{4}+b_{13}\mu l_{4}+b_{14}\theta \Big](1+z)H\Phi'\nonumber\\&&+\Big(b_{5}+b_{14}\mu l_{4}\gamma\Big)\Phi^{k},\label{eq75}
 \end{eqnarray}
 \begin{eqnarray}
 &&(1+z)^{2}H^{2}\mathcal{G}''=-\Big[(1+z)H'+H+\xi\eta-\Big(1-b_{2}\Big)b_{2}+b_{1}b_{13}l_{3}l_{2}-l_{3}b_{6}\Big](1+z)H\mathcal{G}'\nonumber\\ &&-\{\Big(1-b_{2}\Big)\frac{\rho_{m}}{(1+w_{t})\rho_{t}} l_{4}-\xi\Big[\frac{\rho_{m}}{2}-\Upsilon \Big]+b_{13}l_{3}\frac{\rho_{m}}{(1+w_{t})\rho_{t}}l_{2}\}\Delta^{k}_{m}+\Big[\Big(1-b_{2}\Big)b_{14}l_{4}\nonumber\\ &&-\xi\zeta-b_{8} \Big]\Phi^{k}+\Big[\Big(1-b_{2}\Big)b_{13}l_{4}-\xi\Big(2+\gamma\Big)-b^{2}_{13}l_{3}l_{2}-b_{14}l_{3}\Big](1+z)H\Phi'\nonumber\\ &&+\Big[\Big(1-b_{2}\Big)\iota+\xi\Big(1+\varepsilon\Big)+\xi b_{6}+b_{6}b_{13}l_{3}l_{2} \Big]\mathcal{G}^{k}-\Big[\Big(1-b_{2}\Big)\xi+\xi\Big(-\frac{2\theta}{3}+\tau\Big)\nonumber\\ &&-b_{7}b_{13}l_{3}l_{2} -\frac{\rho_{m}}{(1+w_{t})\rho_{t}}l_{3}\Big](1+z)H\Delta'_{m},
 \end{eqnarray}
 \begin{eqnarray}
 &&(1+z)^{2}H^{2}\Phi''=\Big[b_{9}\frac{\rho_{m}}{(1+w_{t})\rho_{t}}l_{4}+b_{11}\Big(\Upsilon-\frac{\rho_{m}}{2}\Big)-b_{12}\frac{\rho_{m}}{(1+w_{t})\rho_{t}}l_{2}\Big]\Delta^{k}_{m}\nonumber\\ &&+\Big(b_{11}\zeta-b_{9}b_{14}l_{3}+b_{12}b_{16}\Big)\Phi^{k}-\Big[b_{9}\iota+b_{11}\Big(1+\varepsilon\Big)+b_{10}+b_{12}b_{17}\Big] \mathcal{G}^{k}\nonumber\\ &&-\Big(b_{2}b_{9}-b_{11}\eta+b_{12}b_{15}\Big)(1+z)H\mathcal{G}'-\Big[(1+z)H'+H+b_{14}l_{1}+\Big(b_{9}b_{13}l_{4}+b_{11}\Big(2+\gamma\Big)\nonumber\\ &&-b_{12}b_{13}l_{4}\Big)\Big](1+z)H\Phi'-\Big[b_{11}\Big(-\frac{2\theta}{3}+\tau\Big)-b_{9}\xi-b_{12}b_{18}-\rho_{m}b_{13}\Big](1+z)H\Delta'_{m}.\label{eq77}
\end{eqnarray}
The eqs. (\ref{eq75})-(\ref{eq77}) are the ones responsible for large scale structure formation for the universe made of dust-scalar field-Gauss-Bonnet energy density perturbations in long wavelength modes. From eqs. (\ref{eq75})-(\ref{eq77}), the matter energy density perturbations couple with the energy density perturbations for the scalar field and Gauss-Bonnet fluids. The numerical solutions of eqs. (\ref{eq75})-(\ref{eq77}) were computed using eqs. (\ref{eq62}--\ref{eq65}) together with $l_{1}-l_{6}$. We chose the initial conditions $\Delta(z)=10^{-5}$ and $\Delta'(z)=0$, together with  $\mathcal{G}(z)=10^{-5}$, $\mathcal{G}'(z)=0$, $\Phi(z)=10^{-5}$ and $\Phi'(z)=0$, and the arbitrary  constants as present in different $f(G)$ models. The numerical solutions are presented in figs. (\ref{Fig9}), (\ref{Fig10}), (\ref{Fig11}) for three different models, respectively. From the plot, the energy density perturbations decay with redshift. The implications of different models on the energy overdensity contrast are presented in table (\ref{table3}) for dust-scalar field-Gauss-Bonnet fluid mixture in the long wavelength mode.
  \begin{table}[ht]
\caption{Illustration of the features of energy overdensity contrast for dust-scalar field-Gauss-Bonnet fluids system in long wavelength mode.}
\centering
\begin{tabular}{c c c c }
\hline
  Models& Ranges of parameters   &  behavior of $\delta(z)$ & Observation \\ [0.5ex]

\hline
 model$1$ &$\alpha_{1}=100$, $\alpha_{2}=0.001$ & decay with increase in $z$& observational supported   \\    \hline

 model$2$ &$0<\beta<\frac{1}{2}$ & decay with increase in $z$ & observational supported  \\   \hline

 model$3$ &$1<n<2$ & grow with increase in $z$ & differ from the $\Lambda$CDM predictions \\[1ex]
\hline
\end{tabular}
\label{table3}
\end{table}

\subsubsection{Perturbations in radiation-scalar field-Gauss-Bonnet fluids dominated Universe\\}
\begin{figure}
  \includegraphics[width=100mm,height=70mm]{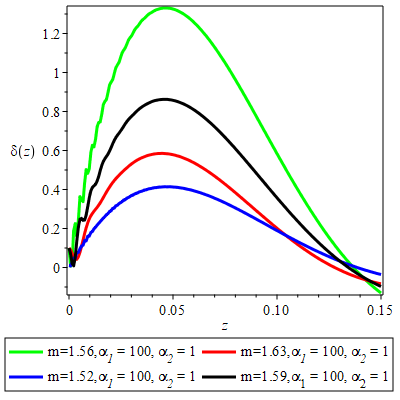}
  \caption{Plot of energy density perturbations versus redshift of Eq. (\ref{eq78}-\ref{eq80}) using model $1$ for different values of $m$ of the radiation-scalar field-Gauss-Bonnet fluids system in a long wavelength mode.}
  \label{Fig12}
  \end{figure}

\begin{figure}
  \includegraphics[width=100mm,height=70mm]{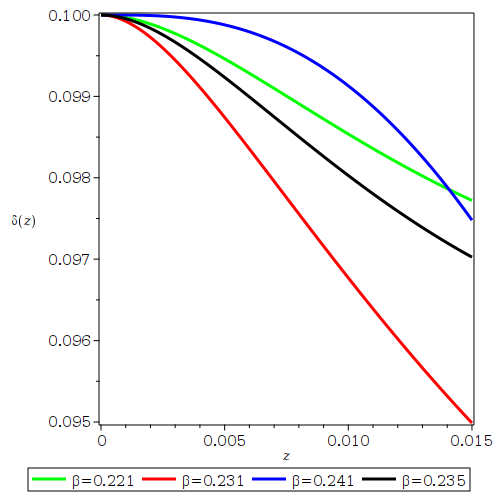}
  \caption{Plot of energy density perturbations versus redshift of Eq. (\ref{eq78}-\ref{eq80}) using model $2$ for different values of $\beta$ of the radiation-scalar field-Gauss-Bonnet fluids system in a long wavelength mode.}
  \label{Fig13}
  \end{figure}

\begin{figure}
  \includegraphics[width=100mm,height=70mm]{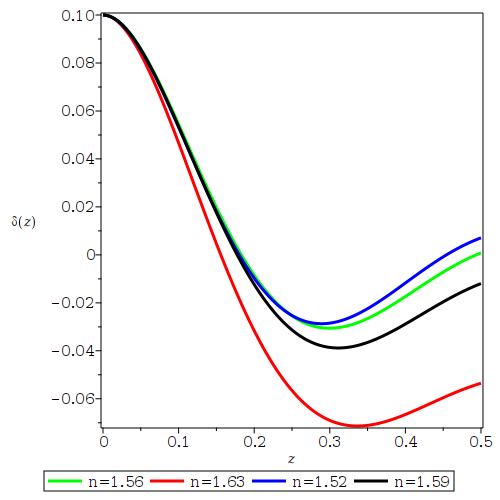}
  \caption{Plot of energy density perturbations versus redshift of Eq. (\ref{eq78}-\ref{eq80}) using model $3$ for different values of $n$ of the radiation-scalar field-Gauss-Bonnet fluids system in a long wavelength mode.}
  \label{Fig14}
  \end{figure}

In the radiation-dominated universe, the equation of state parameter  $w=\frac{1}{3}$ is used  so that eq. (\ref{eq59}) through to eq. (\ref{eq61}) reduce to
\begin{eqnarray}
 &&(1+z)^{2}H^{2}\Delta''_{m}=-\Big[(1+z)H'+H+\frac{16\theta}{9(1+w_{t})\rho_{t}}\rho_{m}+\Big[\frac{4}{3} b_{7}b_{13}\theta l_{2}+\Big(\chi-\frac{4}{3}\Big)\Big(-\frac{2\theta}{3}+\tau\Big)\nonumber\\&&-\mu \xi \Big]\frac{3}{4}\Big](1+z)H\Delta'_{m}+\Big[-b_{3}-\rho_{m}+\Upsilon -\mu\frac{4\rho_{m}}{3(1+w_{t})\rho_{t}}l_{4}-\Big[\frac{4}{3} b_{1}b_{13}\theta l_{2}+\varepsilon+\mu b_{2}\Big]\frac{l_{3}}{4}\nonumber\\&&-\Big(b_{4}+b_{13}\mu l_{4}\Big)\frac{l_{2}}{4}-\Big[\frac{4}{3} b_{7}b_{13}\theta l_{2}+\Big(\chi- \frac{4}{3}\Big)\Big(-\frac{2\theta}{3}+\tau\Big)-\mu \xi \Big]\Big(\frac{\theta}{4}\Big)\Big]\Delta^{k}_{m}\nonumber\\&&+\Big[\frac{4}{3} b_{1}b_{13}\theta l_{2}+\varepsilon+\mu b_{2}-\eta\Big](1+z)H\mathcal{G}'-\Big[\frac{4}{3} b_{6}b_{13}\theta l_{2}+1+\zeta-\mu \iota\Big]\mathcal{G}^{k}\nonumber\\&&+\Big[b_{4}+b_{13}\mu l_{4}+\frac{4}{3} b_{14}\theta \Big](1+z)H\Phi'+\Big(b_{5}+b_{14}\mu l_{4}\gamma\Big)\Phi^{k},\label{eq78}
 \end{eqnarray}
 \begin{eqnarray}
  &&(1+z)^{2}H^{2}\mathcal{G}''=-\Big[(1+z)H'+H+\xi\eta-\Big(1-b_{2}\Big)b_{2}+b_{1}b_{13}l_{3}l_{2}-l_{3}b_{6}\nonumber\\ &&\Big](1+z)H\mathcal{G}'-\{\Big(1-b_{2}\Big)\frac{4\rho_{m}}{3(1+w_{t})\rho_{t}} l_{4}+\Big[\xi\eta-\Big(1-b_{2}\Big)b_{2}+b_{1}b_{13}l_{3}l_{2}\Big]\frac{l_{3}}{4}\nonumber\\&&+\xi\Big[\rho_{m}+\Upsilon \Big]+b_{13}l_{3}\frac{4\rho_{m}}{3(1+w_{t})\rho_{t}}l_{2}+\Big[\Big(1-b_{2}\Big)b_{13}l_{4}+\xi\Big(2+\gamma\Big)-b^{2}_{13}l_{3}l_{2}\Big]\frac{l_{1}}{4}\nonumber\\&&+\Big[\Big(1-b_{2}\Big)\xi+\xi\Big(-\frac{2\theta}{3}+\tau \Big)-b_{7}b_{13}l_{3}l_{2} \Big]\frac{\theta}{4}\}\Delta^{k}_{m}+\Big[\Big(1-b_{2}\Big)b_{14}l_{4}-\xi\zeta-b_{8} \Big]\Phi^{k}\nonumber\\&&+\Big[\Big(1-b_{2}\Big)b_{13}l_{4}-\xi\Big(2+\gamma\Big)-b^{2}_{13}l_{3}l_{2}-b_{14}l_{3}\Big](1+z)H\Phi'+\Big[\Big(1-b_{2}\Big)\iota\nonumber\\&&+\xi\Big(1+\varepsilon\Big)+\xi b_{6}+b_{6}b_{13}l_{3}l_{2} \Big]\mathcal{G}^{k}-\Big[\Big[\Big(1-b_{2}\Big)\xi+\xi\Big(-\frac{2\theta}{3}+\tau\Big)\nonumber\\&&-b_{7}b_{13}l_{3}l_{2} \Big]\Big(\frac{3}{4}\Big)-\frac{4\rho_{m}}{3(1+w_{t})\rho_{t}}l_{3}\Big](1+z)H\Delta'_{m},
 \end{eqnarray}
 \begin{eqnarray}
 &&(1+z)^{2}H^{2}\Phi''=\Big[-\Big(b_{2}b_{9}-b_{11}\eta+b_{12}b_{15}\Big)\frac{l_{3}}{4}\nonumber\\&&+b_{9}\frac{\rho_{m}}{(1+w_{t})\rho_{t}}l_{4}+b_{11}\Big(\Upsilon-\rho_{m}\Big)\nonumber\\&&- b_{12}\frac{4\rho_{m}}{3(1+w_{t})\rho_{t}}l_{2}+\Big(b_{9}b_{13}l_{4}+b_{11}\Big(2+\gamma\Big)-b_{12}b_{13}l_{2}\Big)\frac{ l_{1}}{4}\nonumber\\&&-\Big[b_{11}\Big(-\frac{2\theta}{3}+\tau\Big)-b_{9}\xi-b_{12}b_{18}\Big]\frac{\theta}{4}\Big]\Delta^{k}_{m}+\Big(b_{11}\zeta-b_{9}b_{14}l_{3}+b_{12}b_{16}\Big)\Phi^{k}\nonumber\\&&-\Big[b_{9}\iota+b_{11}\Big(1+\varepsilon\Big)+b_{10}+b_{12}b_{17}\Big] \mathcal{G}^{k}-\Big(b_{2}b_{9}-b_{11}\eta+b_{12}b_{15}\Big)(1+z)H\mathcal{G}'\nonumber\\&&-\Big[\Big((1+z)H'+H\Big)+b_{14}l_{1}+\Big(b_{9}b_{13}l_{4}+b_{11}\Big(2+\gamma\Big)-b_{12}b_{13}l_{4}\Big)\Big](1+z)H\Phi'\nonumber\\&&-\Big[\Big(b_{11}\Big(-\frac{2\theta}{3}+\tau\Big)-b_{9}\xi-b_{12}b_{18}\Big)\Big(\frac{3}{4}\Big)-\frac{4}{3} \rho_{m}b_{13}\Big](1+z)H\Delta'_{m}.\label{eq80}
\end{eqnarray}
The eqs. (\ref{eq78})-(\ref{eq80}) are the energy density perturbations for a radiation-scalar field-Gauss-Bonnet fluid mixture. From eqs. (\ref{eq78})-(\ref{eq80}), the matter energy density perturbations couple with the energy density perturbations for the scalar field and Gauss-Bonnet fluids. The numerical results of eqs. (\ref{eq78})-(\ref{eq80}) were computed using different initial conditions and constants and are presented in figs. (\ref{Fig12}), (\ref{Fig13}), (\ref{Fig14}) for three different $f(G)$ models respectively. From the plots, the energy density perturbations decay with redshift and show some oscillation features. The implications of different models on the energy overdensity contrast are presented in table (\ref{table4}) for radiation-scalar field-Gauss-Bonnet fluid mixture in the long wavelength mode.
  \clearpage
  \begin{table}[ht]
\caption{Illustration of the features of energy overdensity contrast for radiation-scalar field-Gauss-Bonnet fluids system in long wavelength mode.}
\centering
\begin{tabular}{c c c c }
\hline
   Models &Ranges of parameters   &  behavior of $\delta(z)$ & Observation \\ [0.5ex]

\hline
 model$1$ &$\alpha_{1}=100$, $\alpha_{2}=1$ & oscillate then  decay with increase in $z$& observationally supported   \\    \hline

 model$2$ &$0<\beta<\frac{1}{2}$ & decay with increase in  $z$ & observationally supported  \\   \hline

 model$3$ &$1<n<2$ & decay with increase in $z$ & observationally supported  \\[1ex]
\hline
\end{tabular}
\label{table4}
\end{table}
\section{Discussion and Conclusion}\label{sec7}
In the current paper, we present a detailed analysis of cosmological perturbations for the scalar field assisted $f(G)$ gravity. Using the $1+3$ covariant gauge-invariant formalism, we derived the linear covariant evolution equations for a  flat FRW spacetime background. In doing so, we have defined gradient variables for matter fluids, scalar field and Gauss-Bonnet fluids and derived their vector evolution equations. Since our main interest extends to large scale structure formation, we  extracted only scalar parts of the linear evolution equations  using scalar decomposition method, understood to play a key role in structure formation.\\ After getting the scalar perturbation equations, we applied harmonic decomposition method together with the redshift transformation technique to get a set of simplified equations in redshift space responsible for large scale structure formation. For further analysis, we consider two different cases basing on the wavenumber $k$. For short wavelength modes $\frac{k^{2}}{a^{2}H^{2}}\gg1$, we apply the quasi-static approximation to analyze the small scale  fluctuations for a system made of both the mixture of dust-scalar field-Gauss-Bonnet fluids and the mixture of radiation-scalar field-Gauss-Bonnet fluids. For long wavelength modes $\frac{k^{2}}{a^{2}H^{2}}\ll 1$, we solve the whole system of energy density perturbation equations without considering the quasi-static approximation. In both cases, for pedagogical purpose, we  consider a logarthmic scalar field $\phi=\ln(t)$, an exponential potential $V(\phi)=V_{0}e^{-\alpha \phi}$ and a polynomial models $f(G)=\frac{\alpha_{1}}{G^{2}}+\alpha_{2}G^{\frac{1}{2}}$, $f(G)=f_{0}\textbardbl{G}^{\beta},
 f(G)=\frac{c_{1}G^{n}+d_{1}}{c_{2}G^{n}+d_{2}}$  and by using the initial conditions to solve the perturbation equations. For both cases, we get the numerical results which show that the energy density perturbations decay with redshift for different values of $m$. \\ Some of the specific highlights of the present paper include:\\
This work presents the energy overdensity perturbation equations  Eq. (\ref{eq59})-Eq. (\ref{eq61}) in the context of the scalar field assisted $f(G)$ gravity responsible for large scale structure formation.
  The obtained matter energy density perturbation equation couples with the energy density fluctuations resulting from the mixture of scalar field and Gauss-Bonnet fluids in the long wavelength modes while it decouples in the short wavelength modes and GR limits.
  The perturbation equations were solved numerically and the results are presented in figs. (\ref{Fig1})-(\ref{Fig14}). During numerical results computation,  we used different initial conditions, coupling constants ($\alpha_{1}$ and $\alpha_{2}$, $f_{0}$, $c_{1}$, $c_{2}, d_{1}$ and $d_{2}$) together with a range of values of parameters $m$, $\beta$ and $n$ as presented in table (\ref{table1})--(\ref{table4}). For instance, for the case of dust-scalar field-Gauss-Bonnet fluid system in short wave-length, we used $\Delta(z)=10^{-5}$ and $\Delta'(z)=0$, $\alpha_{1}=1$, $\alpha_{2}=0.1$, $k=0$ , for $m=1.73, 1.75, 1.79$, $\beta=0.221, 0.230, 0.241$ and $n=1.73, 1.75, 1.79$ while for radiation-scalar field-Gauss-Bonnet fluid system  $\alpha_{1}=10$, $\alpha_{2}=0.01$, $k=0$ , for $m=1.65, 1.70, 1.78, 1.85$, $\beta=0.211, 0.221, 0.231, 0.240$ and $n=1.88, 1.91, 1.93, 1.95$ were considered. In long wave-length limits, we considered $\Delta(z)=10^{-5}$ and $\Delta'(z)=0$, together with  $\mathcal{G}(z)=10^{-5}$, $\mathcal{G}'(z)=0$, $\Phi(z)=10^{-5}$ and $\Phi'(z)=0$, and $k=1$,$\alpha_{1}=100$, $\alpha_{2}=0.001$, for $m=1.62, 1.65, 1.67$, $\beta=0.221, 0.231, 0.241$ and $n=1.18, 1.25, 1.31$ in case of dust-scalar field-Gauss-Bonnet fluid system while $k=100$, $\alpha_{1}=1$, $\alpha_{2}=1$, for $m=1.52, 1.56, 1.59, 1.63$ $\beta=0.221, 0.231, 0.235, 0.241$ and $n=1.52, 1.56, 1.59, 1.63$ in case of radiation-scalar field-Gauss-Bonnet fluid system give the good results.   From the plots, the energy density perturbations decay with increase in redshift.
   The numerical results are sensible to the parameter $m$, $\beta$ and $n$. As $m$, $\beta$ or $n$ changes, we notice a change in amplitude and the behavior of the curve of energy density perturbations versus redshift. The scalar field assisted $f(G)$ gravity results coincides with GR predictions for $m=1$,$\beta=1$ and $n=1$ for the considered $f(G)$ models.
In general, the combination of scalar field model and $f(G)$ model provides an alternative way of explaining large scale sctructure formation scenarios albeit a variation in the decay in amplitudes of the energy density fluctuations which can help in constrianing the model parameters using different observations. Future works will mainly be based on consideration of different scalar fields with different viable $f(G)$ models in the $1+3$ covariant perturbations to analyse the  implications of scalar field assisted $f(G)$ model on large scale structure formation. It is possible to extend this formulation to different modified gravity theories such as $f(R)$, $f(T)$ to study  cosmological perturbations. For instance, one can take into consideration different scalar fields in the matter lagrangian in $f(R)$, $f(T)$ actions. Another application is that, one can use other types of realistic modified gravity models such as one presented in \cite{nojiri2011unified} to investigate their impact on the large structure formation. The corresponding analysis may lead to some significant qualitative outcomes in comparison with the $\Lambda$CDM. It will be investigated elsewhere.
\section*{Acknoledgements}
The authors thank the anonymous reviewer(s) for the constructive comments towards the signiﬁcant improvement of this manuscript.
AM acknowledges financial support from  the Swedish International Development Agency (SIDA) through International Science Program (ISP) to East African Astrophysics Research Network (EAARN) (grant number  AFRO:05).  AM also acknowledges the hospitality of the Department of Physics of the University of Rwanda, where this work was conceptualized and completed. MM and JN acknowledge the financial support from SIDA through ISP to the University of Rwanda (UR) through Rwanda Astrophysics Space and Climate Science Research Group (RASCRG), Grant number RWA:$01$.
\appendix
\section{Useful Linearised Differential Identities}
For all scalars $f$, vectors $V_a$ and tensors that vanish in the background,
$S_{ab}=S_{\la ab\ra}$, the following linearised identities hold:
\begin{eqnarray}
\left(\D_{\la a}\D_{b\ra}f\right)^{.}&=&\D_{\la a}\D_{b\ra}\dot{f}-\sfrac{2}{3}\Theta\D_{\la a}\D_{b\ra}f+\dot{f}\D_{\la a}A_{b\ra}\label{a0}\;,\\
\ep^{abc}\D_b \D_cf &=& 0 \label{a1}\;, \\
\ep_{cda}\D^{c}\D_{\la b}\D^{d\ra}f&=&\ep_{cda}\D^{c}\D_{( b}\D^{d)}f=\ep_{cda}\D^{c}\D_{ b}\D^{d}f=0\label{a2}\;,\\
\D^2\left(\D_af\right) &=&\D_a\left(\D^2f\right)
+\sfrac{1}{3}\tl{R}\D_a f \label{a4}\;,\\
\left(\D_af\right)^{\rd} &=& \D_a\dot{f}-\sfrac{1}{3}\Theta\D_af+\dot{f}A_a
\label{a14}\;,\\
\left(\D_aS_{b\cdots}\right)^{\rd} &=& \D_a\dot{S}_{b\cdots}
-\sfrac{1}{3}\Theta\D_aS_{b\cdots}
\label{a15}\;,\\
\left(\D^2 f\right)^{\rd} &=& \D^2\dot{f}-\sfrac{2}{3}\Theta\D^2 f
+\dot{f}\D^a A_a \label{a21}\;,\\
\D_{[a}\D_{b]}V_c &=&
-\sfrac{1}{6}\tl{R}V_{[a}h_{b]c} \label{a16}\;,\\
\D_{[a}\D_{b]}S^{cd} &=& -\sfrac{1}{3}\tl{R}S_{[a}{}^{(c}h_{b]}{}^{d)} \label{a17}\;,\\
\D^a\left(\ep_{abc}\D^bV^c\right) &=& 0 \label{a20}\;,\\
\label{divcurl}\D_b\left(\ep^{cd\la a}\D_c S^{b\ra}_d\right) &=& {\ts{1\over2}}\ep^{abc}\D_b \left(\D_d S^d_c\right)\;,\\
\text{curlcurl} V_{a}&=&\D_{a}\left(\D^{b}V_{b}\right)-\D^{2}V_{a}+\sfrac{1}{3}\tl{R}V_{a}\label{curlcurla}\;,
\label{a21}
\end{eqnarray}
\section{Useful relations}
We define
  \begin{eqnarray}
    l_{1}=\dot{\phi},
    l_{2}=\ddot{\phi},
    l_{3}=\dot{G},
    l_{4}=\ddot{G},
    l_{5}=\dddot{G},
    l_{6}=\dddot{\phi}
  \end{eqnarray}
whereas for the quasi-static approximation case, we used
 \begin{eqnarray}
 &&\Big[\Big(1-b_{2}\Big)b_{14}l_{4}-\xi\zeta-b_{8}-b_{14}\xi\frac{k^{2}}{a^{2}} \Big]\Phi^{k}+\Big[\Big(1-b_{2}\Big)\iota+\xi\Big(1+\varepsilon\Big)+\xi b_{6}+b_{6}b_{13}l_{3}l_{2} \Big]\mathcal{G}^{k}\nonumber\\&&=-\{\Big(1-b_{2}\Big)\frac{(1+w_{m})\rho_{m}}{(1+w_{t})\rho_{t}} l_{4}+\Big[\xi\eta-\Big(1-b_{2}\Big)b_{2}+b_{1}b_{13}l_{3}l_{2}\Big]\frac{w_{m}l_{3}}{1+w_{m}}\nonumber\\&&+\xi\Big[-\frac{(1+3w_{m})\rho_{m}}{2}+\Upsilon \Big]+b_{13}l_{3}\frac{(1+w_{m})\rho_{m}}{(1+w_{t})\rho_{t}}l_{2}+\Big[\Big(1-b_{2}\Big)b_{13}l_{4}+\xi\Big(2+\gamma\Big)\nonumber\\&&-b^{2}_{13}l_{3}l_{2}\Big]\Big(\frac{w_{m}l_{1}}{1+w_{m}}\Big)+\Big[\Big(1-b_{2}\Big)\xi+\xi\Big(-\frac{2\theta}{3}+\tau \Big)-b_{7}b_{13}l_{3}l_{2} \Big]\Big(\frac{w_{m}\theta}{1+w_{m}}\Big)\nonumber\\&&+\Big[\xi \frac{(1+w_{m})\rho_{m}}{(1+w_{t})\rho_{t}}+b_{13}\xi\frac{w_{m}l_{1}}{1+w_{m}}-\xi b_{1}\frac{w_{m}l_{3}}{1+w_{m}}+\xi b_{7}\frac{w_{m}\theta}{1+w_{m}}\Big]\frac{k^{2}}{a^{2}}\}\Delta^{k}_{m}-\Big[\Big[\Big(1-b_{2}\Big)\xi\nonumber\\&&+\xi\Big(-\frac{2\theta}{3}+\tau\Big)-b_{7}b_{13}l_{3}l_{2} \Big]\Big(\frac{1}{1+w_{m}}\Big)-\frac{(1+w_{m})\rho_{m}}{(1+w_{t})\rho_{t}}l_{3}-\frac{\xi b_{7}}{1+w_{m}}\frac{k^{2}}{a^{2}}\Big](1+z)H\Delta'_{m},
 \end{eqnarray}
 \begin{eqnarray}
 &&\Big(b_{11}\zeta-b_{9}b_{14}l_{3}+b_{12}b_{16}+b_{11}b_{14}\frac{k^{2}}{a^{2}}\Big)\Phi^{k}-\Big[b_{9}\iota+b_{11}\Big(1+\varepsilon\Big)+b_{10}+b_{12}b_{17}-b_{6}b_{11}\frac{k^{2}}{a^{2}}\Big] \mathcal{G}^{k}\nonumber\\&&=\Big[-\Big(b_{2}b_{9}-b_{11}\eta+b_{12}b_{15}\Big)\Big(\frac{w_{m}l_{3}}{1+w_{m}}\Big)+b_{9}\frac{(1+w_{m})\rho_{m}}{(1+w_{t})\rho_{t}}l_{4}+b_{11}\Big(\Upsilon-\frac{(1+3w_{m})\rho_{m}}{2}\Big)\nonumber\\&&-b_{12}\frac{(1+w_{m})\rho_{m}}{(1+w_{t})\rho_{t}}l_{2}+\Big(b_{9}b_{13}l_{4}+b_{11}\Big(2+\gamma\Big)-b_{12}b_{13}l_{2}\Big)\Big(+\frac{w_{m}l_{1}}{1+w_{m}}\Big)\nonumber\\&&-\Big[b_{11}\Big(-\frac{2\theta}{3}+\tau\Big)-b_{9}\xi-b_{12}b_{18}\Big]\Big(\frac{w_{m}\theta}{1+w_{m}}\Big)+\Big[-b_{11}\frac{(1+w_{m})\rho_{m}}{(1+w_{t})\rho_{t}}-b_{11}b_{13}\Big(\frac{w_{m}l_{1}}{1+w_{m}}\Big)\nonumber\\&&-b_{7}b_{11}\Big(\frac{w_{m}\theta}{1+w_{m}}\Big)+b_{1}b_{11}\Big(\frac{w_{m}l_{3}}{1+w_{m}}\Big)\Big]\frac{k^{2}}{a^{2}}\Big]\Delta^{k}_{m}-\Big[\Big(b_{11}\Big(-\frac{2\theta}{3}+\tau\Big)\nonumber\\&&-b_{9}\xi-b_{12}b_{18}\Big)\Big(\frac{1}{1+w_{m}}\Big)-(1+w_{m})\rho_{m}b_{13}+\frac{b_{7}b_{11}}{1+w_{m}}\frac{k^{2}}{a^{2}}\Big](1+z)H\Delta'_{m}.
\end{eqnarray}
The relations defined below  were used in different parts of the work \begin{eqnarray}
     &&\chi=\frac{(1+w_{m})}{(1+w_{m})\rho_{t}}\Big(8f''H^{2}\ddot{G}+8f''H^{2}\dot{G}^{2}-\frac{Gf''\dot{G}}{3}\Big)\\
     &&\tau=-12f''\dot{G}H+\frac{f''G\dot{G}}{3H^{2}}-8f''\ddot{G}H+8\dot{G}^{2}H-\frac{8f''H\ddot{G}+8f''H\dot{G}^{2}-\frac{Gf''\dot{G}}{3H^{2}}}{3(1+w_{t})\rho_{t}}\Big(-\frac{1}{3}\theta^{2}\nonumber\\&&-\frac{1}{2}(1+3w_{m})\rho_{m}+2\dot{\phi}-2V_{0}e^{-\alpha\phi}-Gf'+f-12f''\dot{G}H^{3}-\frac{f''G\dot{G}}{H}\nonumber\\&&-12H^{2}f''\ddot{G}+12H^{2}f'''\dot{G}^{2}\Big)\\
     &&\Upsilon=-\frac{(1+w_{m})\rho_{m}}{(1+w_{t})\rho_{t}}\Big(-\frac{1}{3}\theta^{2}-\frac{1}{2}(1+3w_{m})\rho_{m}+2\dot{\phi}\nonumber\\&&-2V_{0}e^{-\alpha\phi}-Gf'+f-12f''\dot{G}H^{3}-\frac{f''G\dot{G}}{H}-12H^{2}f''\ddot{G}+12H^{2}f'''\dot{G}^{2}\Big)\\
     &&\gamma=-\frac{\dot{\phi}}{(1+w_{t})\rho_{t}}\Big(-\frac{1}{3}\theta^{2}-\frac{1}{2}(1+3w_{m})\rho_{m}+2\dot{\phi}\nonumber\\&&-2V_{0}e^{-\alpha\phi}-Gf'+f-12f''\dot{G}H^{3}-\frac{f''G\dot{G}}{H}-12H^{2}f''\ddot{G}+12H^{2}f'''\dot{G}^{2}\Big)\\
     &&\zeta=\Big[2-\frac{1}{(1+w_{t})\rho_{t}}\Big(-\frac{1}{3}\theta^{2}-\frac{1}{2}(1+3w_{m})\rho_{m}+2\dot{\phi}\nonumber\\&&-2V_{0}e^{-\alpha\phi}-Gf'+f-12f''\dot{G}H^{3}-\frac{f''G\dot{G}}{H}-12H^{2}f''\ddot{G}\nonumber\\&&+12H^{2}f'''\dot{G}^{2}\Big)\Big]V_{0}\alpha e^{-\alpha \phi}\\
     &&\varepsilon=Gf''+f'+12H^{3}\dot{G}f'''+\frac{f''\dot{G}}{H}+\frac{G\dot{G}f'''}{H}+\frac{12H^{2}f''\dddot{G}}{\dot{G}}+12H^{2}\ddot{G}f'''+12H^{2}\dot{G}^{2}f^{iv}\nonumber\\&&+
 \frac{\frac{1}{2}-\frac{Gf''}{2}+\frac{f'}{2}+\frac{G\dot{G}f'''+\dot{Gf''}}{3H}+\frac{4H^{2}f''\dddot{G}}{\dot{G}}+4H^{2}\ddot{G}f'''+4H^{2}\dot{G}^{2}f'''}{(1+w_{t})\rho_{t}}\Big(-\frac{1}{3}\theta^{2}-\frac{1}{2}(1+3w_{m})\rho_{m}+2\dot{\phi}\nonumber\\&&-2V_{0}e^{-\alpha\phi}-Gf'+f-12f''\dot{G}H^{3}-\frac{f''G\dot{G}}{H}-12H^{2}f''\ddot{G}+12H^{2}f'''\dot{G}^{2}\Big)\\
     &&\eta=12f''H^{3}+\frac{f''G}{H}-24H^{2}f'''\dot{G}+\frac{\frac{Gf''}{3H}+8H^{2}f''\dot{G}}{(1+w_{t})\rho_{t}}\Big(-\frac{1}{3}\theta^{2}-\frac{1}{2}(1+3w_{m})\rho_{m}+2\dot{\phi}\nonumber\\&&-2V_{0}e^{-\alpha\phi}-Gf'+f-12f''\dot{G}H^{3}-\frac{f''G\dot{G}}{H}-12H^{2}f''\ddot{G}+12H^{2}f'''\dot{G}^{2}\Big)\\
 &&\mu=\frac{(1+w_{m})\theta}{(1+w_{t})\rho_{t}}\Big(\frac{1}{2}-\frac{Gf''}{2}+\frac{f'}{2}+\frac{G\dot{G}f'''+\dot{Gf''}}{3H}+\frac{4H^{2}f''\dddot{G}}{\dot{G}}+4H^{2}\ddot{G}f'''+4H^{2}\dot{G}^{2}f'''\Big)
 \end{eqnarray}
\begin{eqnarray}
 &&\nu=\frac{(1+w_{m})\theta}{(1+w_{t})\rho_{t}}\Big(\frac{Gf''}{3H}+8H^{2}f''\dot{G}\Big)\\
 &&\iota=\Big[\frac{\dddot{G}}{\dot{G}}-\frac{\ddot{G}\Big(\frac{1}{2}-\frac{Gf''}{2}+\frac{f'}{2}+\frac{G\dot{G}f'''+\dot{Gf''}}{3H}+\frac{4H^{2}f''\dddot{G}}{\dot{G}}+4H^{2}\ddot{G}f'''+4H^{2}\dot{G}^{2}f'''\Big)}{(1+w_{m})\rho_{t}}\Big],\\
  &&\xi=\Big[\frac{\ddot{G}\Big(8f''H\ddot{G}+8f''H\dot{G}^{2}-\frac{Gf''\dot{G}}{3H^{2}}\Big)}{3(1+w_{t})\rho_{t}}\Big]\\
&& a_{1}=\Big(1-b_{2}\Big)\frac{(1+w_{m})\rho_{m}}{(1+w_{t})\rho_{t}} l_{4}+\Big[\xi\eta-\Big(1-b_{2}\Big)b_{2}+b_{1}b_{13}l_{3}l_{2}\Big]\frac{w_{m}l_{3}}{1+w_{m}}\nonumber\\&&+\xi\Big[-\frac{(1+3w_{m})\rho_{m}}{2}+\Upsilon \Big]+b_{13}l_{3}\frac{(1+w_{m})\rho_{m}}{(1+w_{t})\rho_{t}}l_{2}+\Big[\Big(1-b_{2}\Big)b_{13}l_{4}+\xi\Big(2+\gamma\Big)\nonumber\\&&-b^{2}_{13}l_{3}l_{2}\Big]\Big(\frac{w_{m}l_{1}}{1+w_{m}}\Big)+\Big[\Big(1-b_{2}\Big)\xi+\xi\Big(-\frac{2\theta}{3}+\tau \Big)-b_{7}b_{13}l_{3}l_{2} \Big]\Big(\frac{w_{m}\theta}{1+w_{m}}\Big)\nonumber\\&&+\Big[\xi \frac{(1+w_{m})\rho_{m}}{(1+w_{t})\rho_{t}}+b_{13}\xi\frac{w_{m}l_{1}}{1+w_{m}}-\xi b_{1}\frac{w_{m}l_{3}}{1+w_{m}}+\xi b_{7}\frac{w_{m}\theta}{1+w_{m}}\Big]\frac{k^{2}}{a^{2}}\\
&&a_{2}=\Big[\Big(1-b_{2}\Big)b_{14}l_{4}-\xi\zeta-b_{8}-b_{14}\xi\frac{k^{2}}{a^{2}} \Big] \\
&&a_{3}=\Big[\Big(1-b_{2}\Big)\iota+\xi\Big(1+\varepsilon\Big)+\xi b_{6}+b_{6}b_{13}l_{3}l_{2} \Big]\\
&&a_{4}=-\Big[\Big[\Big(1-b_{2}\Big)\xi\nonumber\\&&+\xi\Big(-\frac{2\theta}{3}+\tau\Big)-b_{7}b_{13}l_{3}l_{2} \Big]\Big(\frac{1}{1+w_{m}}\Big)-\frac{(1+w_{m})\rho_{m}}{(1+w_{t})\rho_{t}}l_{3}-\frac{\xi b_{7}}{1+w_{m}}\frac{k^{2}}{a^{2}}\Big](1+z)H\\
&&a_{5}=\Big[-\Big(b_{2}b_{9}-b_{11}\eta+b_{12}b_{15}\Big)\Big(\frac{w_{m}l_{3}}{1+w_{m}}\Big)+b_{9}\frac{(1+w_{m})\rho_{m}}{(1+w_{t})\rho_{t}}l_{4}+b_{11}\Big(\Upsilon-\frac{(1+3w_{m})\rho_{m}}{2}\Big)\nonumber\\&&-b_{12}\frac{(1+w_{m})\rho_{m}}{(1+w_{t})\rho_{t}}l_{2}+\Big(b_{9}b_{13}l_{4}+b_{11}\Big(2+\gamma\Big)-b_{12}b_{13}l_{2}\Big)\Big(+\frac{w_{m}l_{1}}{1+w_{m}}\Big)\nonumber\\&&-\Big[b_{11}\Big(-\frac{2\theta}{3}+\tau\Big)-b_{9}\xi-b_{12}b_{18}\Big]\Big(\frac{w_{m}\theta}{1+w_{m}}\Big)+\Big[-b_{11}\frac{(1+w_{m})\rho_{m}}{(1+w_{t})\rho_{t}}-b_{11}b_{13}\Big(\frac{w_{m}l_{1}}{1+w_{m}}\Big)\nonumber\\&&-b_{7}b_{11}\Big(\frac{w_{m}\theta}{1+w_{m}}\Big)+b_{1}b_{11}\Big(\frac{w_{m}l_{3}}{1+w_{m}}\Big)\Big]\frac{k^{2}}{a^{2}}\Big]\\
&&a_{6}=\Big(b_{11}\zeta-b_{9}b_{14}l_{3}+b_{12}b_{16}+b_{11}b_{14}\frac{k^{2}}{a^{2}}\Big)\\
&&a_{7}=-\Big[b_{9}\iota+b_{11}\Big(1+\varepsilon\Big)+b_{10}+b_{12}b_{17}-b_{6}b_{11}\frac{k^{2}}{a^{2}}\Big]\\
&&a_{8}=\Big[\Big(b_{11}\Big(-\frac{2\theta}{3}+\tau\Big)\nonumber\\&&-b_{9}\xi-b_{12}b_{18}\Big)\Big(\frac{1}{1+w_{m}}\Big)-(1+w_{m})\rho_{m}b_{13}+\frac{b_{7}b_{11}}{1+w_{m}}\frac{k^{2}}{a^{2}}\Big](1+z)H\\
&&b_{1}=\frac{\Big(\frac{Gf''}{3H}+8H^{2}f''\dot{G}\Big)}{(1+w_{t})\rho_{t}}\\
&&b_{2}=\frac{\ddot{G}\Big(\frac{Gf''}{3H}+8H^{2}f''\dot{G}\Big)}{(1+w_{t})\rho_{t}}\\
&&b_{3}=\frac{(1+w_{m})\theta}{(1+w_{t})\rho_{t}}\dot{\phi}\frac{(1+w_{m})\rho_{m}}{(1+w_{t})\rho_{t}}\ddot{\phi}\\
&&b_{4}=\frac{(1+w_{m})\theta}{(1+w_{t})\rho_{t}}\dot{\phi}\frac{\ddot{\phi}\dot{\phi}}{(1+w_{t})\rho_{t}}\\
&&b_{5}=\frac{(1+w_{m})\theta}{(1+w_{t})\rho_{t}}\dot{\phi}\Big(\frac{\dddot{\phi}}{\dot{\phi}}-\frac{V_{0}\alpha e^{-\alpha \phi}}{(1+w_{t})\rho_{t}}\Big)\\
&&b_{6}=\frac{\frac{1}{2}-\frac{Gf''}{2}+\frac{f'}{2}+\frac{G\dot{G}f'''+\dot{Gf''}}{3H}+\frac{4H^{2}f''\dddot{G}}{\dot{G}}+4H^{2}\ddot{G}f'''+4H^{2}\dot{G}^{2}f'''}{(1+w_{m})\rho_{t}}\\
&&b_{7}=\frac{\Big(8f''H\ddot{G}+8f''H\dot{G}^{2}-\frac{Gf''\dot{G}}{3H^{2}}\Big)}{3(1+w_{t})\rho_{t}},\\
&& b_{8}=\frac{\dot{G}\dot{\phi}}{(1+w_{t})\rho_{t}}\Big[\frac{\dddot{\phi}}{\dot{\phi}}-\frac{V_{0}\alpha e^{-\alpha \phi}\ddot{\phi}}{(1+w_{t})\rho_{t}}\Big]
\end{eqnarray}
\begin{eqnarray}
 && b_{9}=\frac{\dot{\phi}\Big(\frac{Gf''}{3H}+8H^{2}f''\dot{G}\Big)}{(1+w_{t})\rho_{t}}\\
 &&b_{10}=\Big[\frac{\dot{\phi}\Big(\frac{1}{2}-\frac{Gf''}{2}+\frac{f'}{2}+\frac{G\dot{G}f'''+\dot{Gf''}}{3H}+\frac{4H^{2}f''\dddot{G}}{\dot{G}}+4H^{2}\ddot{G}f'''+4H^{2}\dot{G}^{2}f'''\Big)}{(1+w_{m})\rho_{t}}\Big]\\
 &&b_{11}=\Big[\frac{\dot{\phi}\Big(8f''H\ddot{G}+8f''H\dot{G}^{2}-\frac{Gf''\dot{G}}{3H^{2}}\Big)}{3(1+w_{t})\rho_{t}}\Big]\\
 &&b_{12}=\Big(1-\frac{\dot{\phi}^{2}}{(1+w_{t})\rho_{t}}\Big)
\end{eqnarray}
\begin{eqnarray}
 &&b_{13}=\frac{\dot{\phi}}{(1+w_{t})\rho_{t}}\\
 &&b_{14}=-\frac{V_{0}\alpha e^{-\alpha \phi}}{(1+w_{t})\rho_{t}}
\end{eqnarray}
\begin{eqnarray}
 &&b_{15}=-\frac{\ddot{\phi}\Big(\frac{Gf''}{3H}+8H^{2}f''\dot{G}\Big)}{(1+w_{t})\rho_{t}}\\
 &&b_{16}=\Big[\frac{\dddot{\phi}}{\dot{\phi}}-\frac{V_{0}\alpha e^{-\alpha \phi}\ddot{\phi}}{(1+w_{t})\rho_{t}}\Big]\\
 &&b_{17}=\frac{\ddot{\phi}\Big(\frac{1}{2}-\frac{Gf''}{2}+\frac{f'}{2}+\frac{G\dot{G}f'''+\dot{Gf''}}{3H}+\frac{4H^{2}f''\dddot{G}}{\dot{G}}+4H^{2}\ddot{G}f'''+4H^{2}\dot{G}^{2}f'''\Big)}{(1+w_{m})\rho_{t}}\\
 &&b_{18}=\frac{\ddot{\phi}\Big(8f''H\ddot{G}+8f''H\dot{G}^{2}-\frac{Gf''\dot{G}}{3H^{2}}\Big)}{3(1+w_{t})\rho_{t}}\\
 &&E_{1}=\frac{a_{3}a_{7}}{a_{7}a_{2}-a_{6}a_{3}}\Big[\frac{a_{5}}{a_{7}}-\frac{a_{6}a_{1}}{a_{7}a_{2}}-\frac{a_{1}}{a_{2}}\Big] \Big(b_{5}+b_{14}\mu l_{4}\gamma+b_{14}\frac{k^{2}}{a^{2}}\Big)\\
 &&E_{2}=\frac{a_{3} a_{7}}{a_{7} a_{2}-a_{6}a_{3}}\Big[\frac{a_{8}}{a_{7}}-\frac{a_{6}a_{4}}{a_{7}a_{2}}-\frac{a_{4}}{a_{2}}\Big] \Big(b_{5}+b_{14}\mu l_{4}\gamma+b_{14}\frac{k^{2}}{a^{2}}\Big)\\
 &&E_{3}=-\frac{a_{2}a_{7}}{a_{2}a_{7}-a_{3}a_{6}}\Big[\frac{a_{6}a_{1}}{a_{7}a_{2}}-\frac{a_{5}}{a_{7}}\Big]\Big[b_{6}b_{13}(1+w_{m})\theta l_{2}+1+\zeta-\mu \iota-b_{6}\frac{k^{2}}{a^{2}}\Big]\\
 &&E_{4}=-\frac{a_{7} a_{2}}{a_{7} a_{2}-a_{6}a_{3}}\Big[\frac{a_{6}a_{4}}{a_{7}a_{2}}-\frac{a_{8}}{a_{7}}\Big]\Big[b_{6}b_{13}(1+w_{m})\theta l_{2}+1+\zeta-\mu \iota-b_{6}\frac{k^{2}}{a^{2}}\Big]
\end{eqnarray}
\clearpage


\begin{thebibliography}{10}
\expandafter\ifx\csname url\endcsname\relax
  \def\url#1{{\tt #1}}\fi
\expandafter\ifx\csname urlprefix\endcsname\relax\def\urlprefix{URL }\fi
\providecommand{\eprint}[2][]{\url{#2}}

\bibitem{filippenko1998results}
Filippenko A~V and Riess A~G 1998 {\em Physics Reports\/} {\bf 307} 31--44
\bibitem{perlmutter1999measurements}
Perlmutter S, Aldering G, Goldhaber G, Knop R, Nugent P, Castro P~G, Deustua S,
  Fabbro S, Goobar A, Groom D~E {\em et~al.\/} 1999 {\em The Astrophysical
  Journal\/} {\bf 517} 565

\bibitem{riess2004type}
Riess A~G, Strolger L~G, Tonry J, Casertano S, Ferguson H~C, Mobasher B,
  Challis P, Filippenko A~V, Jha S, Li W {\em et~al.\/} 2004 {\em The
  Astrophysical Journal\/} {\bf 607} 665

\bibitem{sherwin2011evidence}
Sherwin B~D, Dunkley J, Das S, Appel J~W, Bond J~R, Carvalho C~S, Devlin M~J,
  D{\"u}nner R, Essinger-Hileman T, Fowler J~W {\em et~al.\/} 2011 {\em
  Physical review letters\/} {\bf 107} 021302

\bibitem{das2011detection}
Das S, Sherwin B~D, Aguirre P, Appel J~W, Bond J~R, Carvalho C~S, Devlin M~J,
  Dunkley J, D{\"u}nner R, Essinger-Hileman T {\em et~al.\/} 2011 {\em Physical
  Review Letters\/} {\bf 107} 021301

\bibitem{eisenstein2005detection}
Eisenstein D~J, Zehavi I, Hogg D~W, Scoccimarro R, Blanton M~R, Nichol R~C,
  Scranton R, Seo H~J, Tegmark M, Zheng Z {\em et~al.\/} 2005 {\em The
  Astrophysical Journal\/} {\bf 633} 560

\bibitem{riess1998observational}
Riess A~G, Filippenko A~V, Challis P, Clocchiatti A, Diercks A, Garnavich P~M,
  Gilliland R~L, Hogan C~J, Jha S, Kirshner R~P {\em et~al.\/} 1998 {\em The
  astronomical journal\/} {\bf 116} 1009

\bibitem{caldwell2009physics}
Caldwell R~R and Kamionkowski M 2009 {\em Annual Review of Nuclear and Particle
  Science\/} {\bf 59} 397--429

\bibitem{silvestri2009approaches}
Silvestri A and Trodden M 2009 {\em Reports on Progress in Physics\/} {\bf 72}
  096901

\bibitem{weinberg2013observational}
Weinberg D~H, Mortonson M~J, Eisenstein D~J, Hirata C, Riess A~G and Rozo E
  2013 {\em Physics reports\/} {\bf 530} 87--255

\bibitem{swart2019unifying}
Swart A~M, Hough R, Sahlu S, Sami H, Tsabone T, Aworka R, Elmardi M and Abebe A
  2019 Unifying dark matter and dark energy in chaplygin gas cosmology {\em
  SAIP conference proceedings\/}

\bibitem{sahlu2019chaplygin}
Sahlu S, Ntahompagaze J, Elmardi M and Abebe A 2019 {\em The European Physical
  Journal C\/} {\bf 79} 1--31

\bibitem{gadbail2022generalized}
Gadbail G~N, Arora S and Sahoo P 2022 {\em Physics of the Dark Universe\/} {\bf
  37} 101074

\bibitem{twagirayezu2023chaplygin}
Twagirayezu F, Ayirwanda A, Munyeshyaka A, Mukeshimana S, Ntahompagaze J and
  Uwimbabazi L~F~R 2023 {\em arXiv preprint arXiv:2305.19711\/}

\bibitem{azizi2017reconstruction}
Azizi T and Naserinia P 2017 {\em The European Physical Journal Plus\/} {\bf
  132} 1--11

\bibitem{hough2021confronting}
Hough R, Sahlu S, Sami H, Elmardi M, Swart A~M and Abebe A 2021 {\em arXiv
  preprint arXiv:2112.11695\/}

\bibitem{saadat2013viscous}
Saadat H and Pourhassan B 2013 {\em International Journal of Theoretical
  Physics\/} {\bf 52} 3712--3720

\bibitem{elmardi2016chaplygin}
Elmardi M, Abebe A and Tekola A 2016 {\em International Journal of Geometric
  Methods in Modern Physics\/} {\bf 13} 1650120

\bibitem{paul2013observational}
Paul B and Thakur P 2013 {\em Journal of Cosmology and Astroparticle Physics\/}
  {\bf 2013} 052

\bibitem{li2007cosmology}
Li B, Barrow J~D and Mota D~F 2007 {\em Physical Review D\/} {\bf 76} 044027

\bibitem{nojiri2008modified}
Nojiri S and Odintsov S~D 2008 {\em Physical Review D\/} {\bf 77} 026007

\bibitem{makarenko2016role}
Makarenko A~N 2016 {\em International Journal of Geometric Methods in Modern
  Physics\/} {\bf 13} 1630006

\bibitem{makarenko2017asymptotic}
Makarenko A~N and Myagky A~N 2017 {\em International Journal of Geometric
  Methods in Modern Physics\/} {\bf 14} 1750148

\bibitem{sami2017inflationary}
Sami H, Ntahompagaze J and Abebe A 2017 {\em Universe\/} {\bf 3} 73

\bibitem{nojiri2017modified}
Nojiri S, Odintsov S and Oikonomou V 2017 {\em Physics Reports\/} {\bf 692}
  1--104

\bibitem{ntahompagaze2017f}
Ntahompagaze J, Abebe A and Mbonye M 2017 {\em International Journal of
  Geometric Methods in Modern Physics\/} {\bf 14} 1750107

\bibitem{odintsov2017unification}
Odintsov S, Oikonomou V and Sebastiani L 2017 {\em Nuclear Physics B\/} {\bf
  923} 608--632

\bibitem{sami2018reconstructing}
Sami H, Namane N, Ntahompagaze J, Elmardi M and Abebe A 2018 {\em International
  Journal of Geometric Methods in Modern Physics\/} {\bf 15} 1850027

\bibitem{capozziello2019cosmological}
Capozziello S, Mantica C~A and Molinari L~G 2019 {\em International Journal of
  Geometric Methods in Modern Physics\/} {\bf 16} 1950008

\bibitem{odintsov2019unification}
Odintsov S and Oikonomou V 2019 {\em Physical Review D\/} {\bf 99} 104070

\bibitem{nojiri2020unifying}
Nojiri S, Odintsov S~D, Oikonomou V~K and Paul T 2020 {\em Physical Review D\/}
  {\bf 102} 023540

\bibitem{odintsov2020geometric}
Odintsov S~D and Oikonomou V~K 2020 {\em Physical Review D\/} {\bf 101} 044009
\bibitem{venikoudis2022late}
Venikoudis S, Fasoulakos K and Fronimos F 2022 {\em International Journal of
  Modern Physics D\/} {\bf 31} 2250038
\bibitem{kobayashi2009can}
Kobayashi Tsutomu and Maeda Kei-ichi 2009 {\em Physical Review D\/} {\bf 79} 024009
\bibitem{babichev2010relativistic}
Babichev Eugeny and Langlois David 2010 {\em Physical Review D\/} {\bf 79} 124051
\bibitem{bamba2008future}
Bamba Kazuharu, Nojiri Shin’ichi and Odintsov Sergei D 2008 {\em Journal of Cosmology and Astroparticle Physics\/} {\bf 10} 045
\bibitem{bamba2010finite}
Bamba Kazuharu, Odintsov Sergei D, Sebastiani Lorenzo and Zerbini Sergio 2010 {\em The European Physical Journal C\/} {\bf 67} 295--310
\bibitem{dolgov2003can}
Dolgov Alexander D and Kawasaki Masahiro 2003 {\em Physics Letters B\/} {\bf 573} 1--4
\bibitem{nojiri2006dark}
Nojiri S, Odintsov S~D and Gorbunova O 2006 {\em Journal of Physics A:
  Mathematical and General\/} {\bf 39} 6627

  \bibitem{nojiri2008inflation}
Nojiri Shin'ichi, Odintsov Sergei D and Tretyakov Petr V 2008 {\em Progress of Theoretical Physics Supplement\/} {\bf 172} 81--89
\bibitem{nojiri2005modified}
Nojiri S and Odintsov S~D 2005 {\em Physics Letters B\/} {\bf 631} 1--6
\bibitem{nojiri2011unified}
Nojiri Shin'ichi and Odintsov Sergei D 2011 {\em Physics Reports\/} {\bf 505} 59--144






\bibitem{nojiri2005gauss}
Nojiri Shin’ichi, Odintsov Sergei D and Sasaki Misao 2005 {\em Physical Review D\/} {\bf 71} 123509
\bibitem{calcagni2005dark}
Calcagni, Gianluca and Tsujikawa, Shinji and Sami, M 2005 {\em Classical and Quantum Gravity\/} {\bf 22} 3977
\bibitem{cognola2007string}
Cognola Guido, Elizalde Emilio, Nojiri Shin’ichi, Odintsov Sergei D and Zerbini Sergio 2007 {\em Physical Review D\/} {\bf 75} 086002
\bibitem{nojiri2007dark}
Nojiri Shin'ichi, Odintsov Sergei D and Tretyakov Petr V 2007 {\em Physics Letters B\/} {\bf 651} 224--231
\bibitem{odintsov2020unification}
Odintsov S~D, Oikonomou V~K, Fronimos F and Fasoulakos K 2020 {\em Physical
  Review D\/} {\bf 102} 104042
\bibitem{oikonomou2016gauss}
Oikonomou V 2016 {\em Astrophysics and Space Science\/} {\bf 361} 211
\bibitem{oikonomou2020non}
Oikonomou V and Fronimos F 2020 {\em The European Physical Journal Plus\/} {\bf
  135} 917

\bibitem{odintsov2020rectifying}
Odintsov S~D, Oikonomou V~K and Fronimos F 2020 {\em Nuclear Physics B\/} {\bf
  958} 115135

\bibitem{oikonomou2021refined}
Oikonomou V 2021 {\em Classical and Quantum Gravity\/} {\bf 38} 195025

\bibitem{odintsov2021canonical}
Odintsov S~D, Oikonomou V and Fronimos F 2021 {\em Annals of Physics\/} {\bf
  424} 168359
\bibitem{fronimos2021inflation}
Fronimos F and Venikoudis S 2021 {\em International Journal of Modern Physics
  A\/} {\bf 36} 2150229
\bibitem{bardeen1980gauge}
Bardeen J~M 1980 {\em Physical Review D\/} {\bf 22} 1882

\bibitem{dunsby1992covariant}
Dunsby P~K, Bruni M and Ellis G~F 1992 {\em Astrophysical Journal, Part 1 (ISSN
  0004-637X), vol. 395, no. 1, p. 54-74.\/} {\bf 395} 54--74

\bibitem{perico2017running}
Perico E and Tamayo D 2017 {\em Journal of Cosmology and Astroparticle
  Physics\/} {\bf 2017} 026

\bibitem{borges2020growth}
Borges H~A and Wands D 2020 {\em Physical Review D\/} {\bf 101} 103519
\bibitem{sharma2021growth}
Sharma M~K and Sur S 2021 Growth of matter perturbations in an interacting dark
  energy scenario emerging from metric-scalar-torsion couplings {\em Physical
  Sciences Forum\/} vol~2 (MDPI) p~51
\bibitem{ellis1989covariant}
Ellis G~F and Bruni M 1989 {\em Physical Review D\/} {\bf 40} 1804

\bibitem{dunsby1992cosmological}
Dunsby P, Bruni M and Ellis G 1992 {\em Astrophys. J\/} {\bf 395} 34

\bibitem{bruni1992gauge}
Bruni M, Ellis G~F and Dunsby P~K 1992 {\em Classical and Quantum Gravity\/}
  {\bf 9} 921

\bibitem{sahlu2020scalar}
Sahlu S, Ntahompagaze J, Abebe A, de~la Cruz-Dombriz {\'A} and Mota D~F 2020
  {\em The European Physical Journal C\/} {\bf 80} 422

\bibitem{ntahompagaze2020multifluid}
Ntahompagaze J, Sahlu S, Abebe A and Mbonye M~R 2020 {\em International Journal
  of Modern Physics D\/} {\bf 29} 2050120
\bibitem{maartens1998covariant}
Maartens R 1998 {\em Physical Review D\/} {\bf 58} 124006

\bibitem{sami2021perturbations}
Sami H and Abebe A 2021 {\em International Journal of Geometric Methods in
  Modern Physics\/} {\bf 18} 2150158
\bibitem{borges2008evolution}
Borges H, Carneiro S, Fabris J and Pigozzo C 2008 {\em Physical Review D\/}
  {\bf 77} 043513
\bibitem{munyeshyaka2023multifluid}
Munyeshyaka A, Ayirwanda A, Twagirayezu F, Murorunkwere B and Ntahompagaze J
  2023 {\em International Journal of Geometric Methods in Modern Physics\/}
  {\bf 20} 2350031

\bibitem{munyeshyaka2021cosmological}
Munyeshyaka A, Ntahompagaze J and Mutabazi T 2021 {\em International Journal of
  Modern Physics D\/} {\bf 30} 2150053

\bibitem{munyeshyaka20231+}
Munyeshyaka A, Ntahompagaze J, Mutabazi T, Mbonye M {\em et~al.\/} 2023 {\em
  arXiv preprint arXiv:2305.01331\/}

\bibitem{munyeshyaka2023perturbations}
Munyeshyaka A, Ntahompagaze J, Mutabazi T, Mbonye M~R, Ayirwanda A, Twagirayezu
  F and Abebe A 2023 {\em International Journal of Geometric Methods in Modern
  Physics\/} {\bf 20} 2350047
\bibitem{copeland2006dynamics}
Copeland E~J, Sami M and Tsujikawa S 2006 {\em International Journal of Modern
  Physics D\/} {\bf 15} 1753--1935
\bibitem{odintsov2021late}
Odintsov S~D, Oikonomou V~K and Fronimos F 2021 {\em Classical and Quantum
  Gravity\/} {\bf 38} 075009
\bibitem{gourgoulhon20073+}
Gourgoulhon E 2007 {\em arXiv preprint gr-qc/0703035\/}
\bibitem{park2018covariant}
Park C 2018 {\em arXiv preprint arXiv:1810.06293\/}
\bibitem{boehmer2008dynamics}
Boehmer C~G, Caldera-Cabral G, Lazkoz R and Maartens R 2008 {\em Physical
  Review D\/} {\bf 78} 023505
\bibitem{abebe2012covariant}
Abebe A, Abdelwahab M, De~la Cruz-Dombriz A and Dunsby P~K 2012 {\em Classical
  and quantum gravity\/} {\bf 29} 135011

\bibitem{carloni2006gauge}
Carloni S, Dunsby P~K and Rubano C 2006 {\em Physical Review D\/} {\bf 74}
  123513

\bibitem{abebe2015breaking}
Abebe A 2015 {\em International Journal of Modern Physics D\/} {\bf 24} 1550053
\bibitem{sahlu2020perturbations}
Sahlu S, Sami H, Swart A~M, Tsabone T, Elmardi M and Abebe A 2020 {\em arXiv
  preprint arXiv:2010.05173\/}
\bibitem{bamba2017energy}
Bamba Kazuharu, Ilyas M, Bhatti MZ and Yousaf Z 2017 {\em General Relativity and Gravitation\/} {\bf 49}
  1--17
\bibitem{de2009construction}
De Felice Antonio and Tsujikawa Shinji 2009 {\em Physics Letters B\/} {\bf 675}
  1--8
\bibitem{dunsby1991gauge}
Dunsby P~K 1991 {\em Classical and Quantum Gravity\/} {\bf 8} 1785
\end{thebibliography}
\bibliographystyle{iopart-num}
\providecommand{\newblock}{}

 \noindent
{\color{blue} \rule{\linewidth}{1mm} }
  \end{document}